# Exploring the added value of pretherapeutic MR descriptors in predicting breast cancer pathologic complete response to neoadjuvant chemotherapy

C Malhaire[1,2] MD, F Selhane[3] MD, MJ Saint-Martin[2] PhD student, V Cockenpot[4] MD, P Akl[5] MD, E Laas[6] MD, A Bellesoeur[7] MD, C Ala Eddine [1] MD, M Bereby-Kahane [1] MD, J Manceau [1] MD, D Sebbag-Sfez [1] MD, JY Pierga[7] MD, PhD, F Reyal[6] MD, PhD, A Vincent-Salomon[8] MD, PhD, H Brisse [1] MD, PhD, F Frouin [2] PhD

[1] Institut Curie, Department of Medical Imaging, PSL research university, 26 rue d'Ulm, 75005 Paris, France

[2] Institut Curie, Research Center, U1288-LITO, Inserm, Paris-Saclay University, 91401 Orsay, France

[3] Gustave Roussy, Department of Imaging, Paris-Saclay University, 94805 Villejuif, France

[4] Centre Léon Bérard, Pathology Unit, 28 rue Laennec, 69008 Lyon, France

[5] HCL, Radiologie du groupement hospitalier Est, Women Imaging Unit, 3 Quai des Célestins, 69002 Lyon, France

[6] Institut Curie, Department of Surgical Oncology, 26 rue d'Ulm, 75005 Paris, France

[7] Institut Curie, Department of Medical Oncology, 26 rue d'Ulm, 75005 Paris, France

[8] Institut Curie, Department of Pathology, 26 rue d'Ulm, 75005 Paris, France

**List of authors :**

Caroline Malhaire[1,2] MD, MSc, PSL research university

Fatine Selhane [1] MD , University of Paris

Marie-Judith Saint-Martin[2] PhD student, Paris-Saclay University

Vincent Cockenpot[3] MD, PSL research university

Pia Akl [1,2,4] MD, MSc, PSL research university

Enora Laas[5] MD, PSL research university

Audrey Bellesoeur[6] MD, PSL research university

Catherine Ala Eddine [1] MD, PSL research university

Mélodie Bereby-Kahane [1] MD, PSL research university

Julie Manceau [1] MD, PSL research university

Delphine Sebbag-Sfez [1] MD, PSL research university

Jean-Yves Pierga[6] MD, PhD, Paris Descartes University

Fabien Reyal[5] MD, PhD, Paris Descartes University

Anne Vincent-Salomon[3] MD, PSL research university

Hervé Brisse [1] MD, PhD, PSL research university

Frédérique Frouin [2] PhD, Paris-Saclay University

**Institution :**

Institut Curie, 26 rue d'Ulm, 75005 Paris, France

**Corresponding Author:**

Caroline Malhaire,

Department of Medical Imaging, Institut Curie, 26 rue d'Ulm, 75005 Paris, France

Email : caroline.malhaire@curie.fr

Tel : 0033+ 1 44 32 42 81

**Manuscript type:** Original Research

**Disclosure:** The authors have nothing to declare in relation to the content of this article

# Exploring the added value of pretherapeutic MR descriptors in predicting breast cancer pathologic complete response to neoadjuvant chemotherapy

**Manuscript type:** Original research

**Abbreviations**

BES: Breast Edema Score

BC: Breast Cancer

BI-RADS: Breast Imaging & Reporting Data System

HER2: Human Epidermal growth factor Receptor 2

NAC: Neoadjuvant Chemotherapy

NME: NonMass Enhancement

pCR: pathologic Complete Response

RCB: Residual Cancer Burden

TIL: Tumor-Infiltrating Lymphocyte

TN: Triple Negative

**Abstract**

Objectives: To evaluate the association between pretreatment MRI descriptors and breast cancer (BC) pathological complete response (pCR) to neoadjuvant chemotherapy (NAC).

Materials & Methods: Patients with BC treated by NAC with a breast MRI between 2016 and 2020 were included in this retrospective observational single-center study. MR studies were described using the standardized BI-RADS and breast edema score on T2-weighted MRI. Univariable and multivariable logistic regression analyses were performed to assess variables association with pCR according to residual cancer burden. Random forest classifiers were trained to predict pCR on a random split including 70% of the database and were validated on the remaining cases.

Results: Among 129 BC, 59 (46%) achieved pCR after NAC (luminal (n=7/37, 19%), triple negative (TN) (n=30/55, 55%), HER2+ (n=22/37, 59%)). Clinical and biological items associated with pCR were BC subtype ($p<0.001$), T stage 0/I/II ($p=0.008$), higher Ki67 ($p=0.005$) and higher tumor-infiltrating lymphocytes levels ($p=0.016$). Univariate analysis showed that the following MRI features, oval or round shape ($p=0.047$), unifocality ($p=0.026$), non-spiculated margins ($p=0.018$), no associated non-mass enhancement (NME) ($p = 0.024$) and a lower MRI size ($p = 0.031$) were significantly associated with pCR. Unifocality and non-spiculated margins remained independently associated with pCR at multivariable analysis. Adding significant MRI features to clinicobiological variables in random forest classifiers significantly increased sensitivity (0.67 versus 0.62), specificity (0.69 versus 0.67) and precision (0.71 versus 0.67) for pCR prediction.

Conclusion: Non-spiculated margins and unifocality are independently associated with pCR and can increase models performance to predict BC response to NAC.

Clinical Relevance Statement: A multimodal approach integrating pretreatment MRI features with clinicobiological predictors, including TILs, could be employed to develop machine learning models for identifying patients at risk of non-response. This may enable consideration of alternative therapeutic strategies to optimize treatment outcomes

**Key points:**

. Unifocality and non-spiculated margins are independently associated with pCR at multivariable logistic regression analysis.

- Breast Edema Score is associated with MR tumor size and TILs expression, not only in TN BC as previously reported, but also in luminal BC.

. Adding significant MRI features to clinicobiological variables in machine learning classifiers significantly increased sensitivity, specificity, and precision for pCR prediction.

**Introduction**

Neoadjuvant chemotherapy (NAC) is indicated in locally advanced breast cancer (BC), to downstage large tumors and allow breast-conserving surgery after BC response, producing no significant differences in terms of overall and disease-free survival compared to surgery followed by adjuvant chemotherapy [1]. NAC also provides valuable *in vivo* information regarding tumor response of aggressive early-stage BC. Complete breast and axillary pathological response is associated with improved long-term survival, particularly for triple negative and human epidermal growth factor receptor (HER) 2-positive tumors [2–4].

MRI is the modality of choice for pretreatment staging and tumor response monitoring during NAC and has been demonstrated to improve the prediction of tumor response after NAC using morphological and functional parameters such as diffusion-weighted imaging [5, 6]. The Breast Imaging Reporting and Data System (BI-RADS) MRI lexicon has been built and updated to standardize breast MRI lesion description and reporting, using specific terms that can be described and compared using the wide range of available MR hardware and software [7]. Several BC pretreatment MRI features from the BI-RADS lexicon have been selectively studied and found to be associated with tumor response after NAC, such as round or oval shape, homogeneous enhancement, washout enhancement in populations of unspecified BC subtypes and triple negative BC [8, 9], with, however, little information regarding luminal and HER2-positive carcinomas.

In addition, T2-weighted MRI features, such as the absence of a peritumoral edema, are associated with pathologic complete response (pCR) and recurrence free-survival in TN BC [8, 10]. Cheon *et al.* demonstrated that the presence of peritumoral edema was associated with the characteristics of biologically aggressive tumors [11], and was one of the most relevant features for the prediction of response to NAC by machine learning classifiers using multiparametric MRI [12]. The peritumoral edema improvement under neoadjuvant chemotherapy may also be a predictor of a more favorable prognosis [13]. Peritumoral edema on MRI scans following NAC was linked to a worse outcome in a large cohort of patients with luminal breast cancer treated with neoadjuvant chemotherapy [14]. The presence of prepectoral edema is also a strong prognostic indicator for lymphatic invasion and has been reported to be associated with high tumor grading and chemoresistance [15, 16].

Hence, Harada *et al.* proposed a breast edema score (BES) recording peritumoral edema, prepectoral edema, and subcutaneous edema on T2-weighted MRI scans. These authors showed that patients treated by NAC for all types of BC with MRI-detected inflammatory BC had poorer progression-free survival, compared to patients with no edema [17], but did not find any correlation of intramammary edema and intratumoral necrosis with pCR in a cohort of TN BC [18].

Our objective was therefore to study the association between pCR after NAC and pretreatment standardized MRI features comprehensively described using the BI-RADS lexicon and BES in a population of TN, HER2-positive and luminal BC, as well as the association of these MRI descriptors with known clinical and biologic predictive biomarkers of pCR.

**Materials and Methods**

**Study design and study population:**

Our institutional review board approved this retrospective analysis of breast MRI that were consecutively performed in patients previously enrolled in a prospective trial (NCT02834494), after giving their written informed consent. Patients were not included if previously treated for ipsilateral breast cancer, pregnant or breast-feeding, or with breast implant(s).

**MRI protocol**

All MRI scans were performed in the prone position with a dedicated breast coil. The pretreatment MRI scans of 91 patients were performed using a 1.5 T magnet, MAGNETOM Aera (Siemens), with an 18-channel coil (coil 1) or an 16-channel coil (Sentinelle; Philips Medical Systems, coil 2) or a 1.5 T magnet Optima MR450w (General Electric Healthcare) with an 8-channel coil (coil 3), using parameters presented in Table 1. Before dynamic contrast-enhanced sequences, an intravenous injection of gadolinium-based contrast material (gadoterate meglumine, Dotarem 0.5mmol/mL, Guerbet Healthcare; 0.2ml per kilogram of body weight) was performed using a power-injector, followed by a 20mL saline solution flush. The MRI performed outside our institution in various centers for 38 patients were reviewed to control image quality and included at least T2-weighted axial sequences with fat suppression and T1 fat-suppressed dynamic contrast-enhanced sequences and subtracted images in compliance with American College of Radiology recommendations with a slice thickness ≤ 3 mm and an in-plane resolution < 1 mm [19].

**Image analysis**

The MRI were reviewed by a radiology resident (FS, 4th year of radiology specialization, 1 year of experience in breast MRI) and an experienced breast radiologist (CM,15 years of experience in breast MRI) blinded to the pathology results. Two readers (CM & FS) first independently reviewed all MRI scans to assess interobserver agreement for the description of MRI features. MRI features were described according to the BI-RADS lexicon [19]. Subsequently, the results of a consensus reading were used for statistical analyses.

Intratumoral high T2 signal intensity was defined as previously published as intensity equal to or greater than that of water or vessels [8]. Peritumoral, prepectoral and subcutaneous edema were analyzed on T2-weighted images to record the BES classifying breast edema into 4 groups illustrated in Figure S1 [18]: BES1 no edema, BES2 peritumoral edema, BES3 prepectoral edema, BES4 subcutaneous edema, whether clinically evident or not.

Kinetic analysis was performed with time-intensity curves performed by placing a region-of-interest on the most enhancing area within the index mass.

Multifocality was defined as the presence of additional malignant sites within the ipsilateral breast. A non-mass enhancement (NME) associated with the index tumor was recorded when present.

MRI sizes of the index mass and the maximal extension of multifocal lesions and/or NME were measured along the longest axis, in the axial, coronal or sagittal plane on the first subtracted axial images of dynamic contrast-enhanced MRI sequences.

**Clinical data and biological parameters**

Clinical variables collected from medical records were: age, BMI, menopausal status, clinical stage according to the TNM staging system [20].

Pathological variables collected from pretreatment biopsies were: histological type, histological grade [21], Ki67 index, TILs assessment according to international recommendations [22]. We defined two groups of low and high levels using cut-offs of 30% for TILs and 25% for Ki67, based on previously described thresholds [23–25]. Estrogen and progesterone receptor positivity was defined by a positive staining > 10%. HER2-positive BC was defined by HER2 overexpression by immunohistochemistry staining and /or fluorescence in situ hybridization [26]. HR-positive/HER2-negative BC was defined as luminal BC, HR-negative/HER2-negative BC was defined as TN BC and HER2-positive BC, irrespective of HR staining, was defined as HER2-positive BC.

Histological tumor response was assessed on post-NAC surgical specimens according to residual cancer burden (RCB), a standardized measure of pathological response that accurately correlates with prognosis in patients with BC treated with NAC [27]. A pathologic complete response (RCB0) according to the RCB is defined by the absence of residual invasive lesions in the breast and axilla, whether or not carcinoma in situ lesions remain after neoadjuvant therapy. The primary endpoint was pCR defined as RCB0 (no invasive residual disease), as opposed to non pCR, including RCB I (minimal residual disease), RCB II (moderate residual disease) and RCB III (extensive residual disease) classes.

**Statistical analysis**

Statistical analyses were performed with R software (version R-4.1.0). All statistical tests were two-sided, and p values less than 0.05 were considered statistically significant. Continuous data are reported as medians +/- interquartile range. We compared the various characteristics between groups by using Wilcoxon rank sum test, Pearson's Chi-squared test or Fisher's exact test, as appropriate. Bonferroni–Holm correction was applied for pairwise multiple comparisons of index tumor MR size distribution between BES classes.

The Cohen κ statistic was used to evaluate interreader agreement, using linear weighted κ for ordinal variables.

Univariate logistic regression analyses were performed to assess variables association with pCR. Multivariable regression analysis was performed using forward selection of variables, and odds-ratios (OR) and 95% confidence intervals (95% CI) for pCR were calculated.

Two random forest models were built: the first one using significant biological and clinical variables, the second adding significant MRI features. For the first random forest classifier, BC subtype, T stage, TILs, Ki67 were used. Maximal MR size, shape, margins and associated NME were added to these clinical and biological features in the second classifier. The classifiers were trained on 70% of cases, and validated on the 30% remaining cases, the random split being repeated 100 times. Sensitivity, specificity, precision, and area under the curve (AUC) were computed and compared using Wilcoxon signed-rank and Delong tests.

## Results

**Patient characteristics**

Our study included 129 patients (mean age, 47 years; range, 24-73 years). No patient had a prior history of treatment for invasive breast cancer. Patients were treated by NAC for BC between 2016 and 2020. Eleven patients were excluded because of MRI technical failure, missing data or if the MRI was performed after the first chemotherapy cycle (Figure 1). Neoadjuvant chemotherapy anthracycline/taxane-based regimens consisted of 4 cycles of anthracycline (AC dose dense or EC) followed by weekly paclitaxel or every-3-week administration of docetaxel. Patients included in the study received an anthracycline-taxane containing regimen (n=126), taxane-only (n=1), anthracycline-only (n=2), with the adjunction of other chemotherapy (n=6). Patients with HER2-positive BC received trastuzumab in addition to chemotherapy.

Clinical and pathological characteristics are presented by pCR status in Table 2, and by BC subtype in Table S1. After NAC completion, pCR was observed in 59 patients (46%). We found no significant differences in distribution of age, BMI, menopausal status and tumor grade between pCR and non-pCR groups.

Response rates differed significantly according to BC subtype ($p<0.001$): the vast majority of luminal BC did not achieve pCR (n=30, 81%), whereas TN and HER2-positive BC were less likely to show residual disease (Table 2). Patients with clinically advanced disease (T stage III/IV) were less likely to achieve pCR ($p=0.008$).

Ki67 (median 60% vs 40%, $p=0.005$) and TILs (median 30% vs 15%, $p=0.016$) levels were significantly higher in the pCR group than in the non-pCR group. The highest Ki67 and TILs levels were observed in the TN BC group, in which grade 3 BC were also more frequent ($p=0.004$).

**Inter-reader agreement**

Inter-reader agreement for the assessment of MRI features was excellent for internal enhancement pattern, intra-tumoral high signal intensity on T2-weighted images, associated NME, prepectoral edema, subcutaneous edema, peritumoral edema, shape and margins (κ=0.81-0.95) (Table S2).

**Association of MRI features with tumor response**

All tumors included presented mass-like enhancements on MRI with a mean index tumor size of 26 mm (range, 12-85 mm) on MRI. On univariate logistic regression analyses (Table 3), tumors achieving pCR were more likely to show an oval or round shape (OR, 2.25, *p*=0.047), and a unifocal distribution (OR, 2.73, *p*=0.026). Tumors with spiculated margins (OR, 0.42, *p*=0.018), associated NME (OR, 0.33, *p* = 0.024) and a larger maximal MRI size (OR, 0.98, *p* = 0.031) were less likely to achieve pCR. Other features such as background parenchymal enhancement, breast composition, presence of an intra-tumoral high signal intensity on T2-weighted images, early internal enhancement and delayed enhancement patterns were not significantly associated with pCR in the whole study population. BES distribution was not significantly different between pCR and non-pCR groups but was significantly associated with MR index tumor size, as tumor size increased with BES score, in the whole population (*p*<0.001), as well as in the luminal (*p*=0.050) and TN (*p*=0.050) BC groups (Figure S2). The presence of subcutaneous edema was also correlated with inferior TILS levels (Figure S3).

Multivariable logistic analysis showed that non-spiculated tumor margins (*p*=0.045) and the absence of multifocality (*p*=0.038) remained independently associated with pCR after adjusting on BC subtype, TILs levels, T Stage and BMI (Table S3). The forest plot displaying OR and their confidence intervals is shown in Figure S4.

The results of a multiple component analysis, is presented in Figure S5 to display associations between the dataset variables. This type of factorial analysis can be viewed as a counterpart of principal component analysis for categorical data. The main guidelines when reading a graph from a multiple component analysis are the following: a) when two lines are globally parallel, the two variables are correlated, b) If the lines intersect at right angles, the two variables are independent, c) when the points of two modalities of two distinct variables are close together, the subjects belonging to one of the modalities share traits with the individuals belonging to the modality of the other variable. The plot shows how spiculated margins are associated with an incomplete response, low KI67 and low TILs expressions, while other variables associated with an incomplete response such as multifocality, an irregular shape are more related to luminal BC subtype than to TILs or KI67 levels.

**pCR prediction**

In the 100 validation sets, adding MRI features to clinical and biologic characteristics improved median sensitivity from 0.62 [95% CI 0.60-0.64] to 0.67 [95% CI 0.64-0.67] ($p$=0.006), median specificity from 0.67 [95% CI 0.62-0.68] to 0.69 [95% CI 0.67-0.71] ($p$=0.007), median precision from 0.67 [95% CI 0.66-0.69] to 0.71 [95% CI 0.70-0.73] ($p$=0.001) and median AUC from 0.63 [95% CI 0.62-0.65] to 0.66 [95% CI 0.66-0.68] ($p$=0.016). In 14 out of 100 iterations, the Delong test demonstrated statistically significant results in favor of a higher AUC for random forest models after incorporating MRI variables. The six main variables selected by random forest classifiers were, by decreasing value of importance: maximal MR size, Ki67, TILs, BC subtype, margins and shape.

**Breast Cancer Subtypes**

Figure 2 presents the tumor response according to RCB classes by BC subtype for each significant MRI features at univariate logistic regression analysis. MRI features association with pCR is presented by BC subtype in Table S4.

Distributions of MRI features and corresponding pCR rates by breast cancer subtype are depicted on mosaic plots in Figure 3.

As illustrated in Figure 4, 95% of luminal BC with spiculated margins did not achieve pCR after NAC, as compared to 62% of tumors with circumscribed/irregular margins ($p$=0.029). None of the luminal tumors with multifocal distribution (n=10) achieved pCR. Lower pCR rates were significantly associated with increasing maximal and index tumor MR sizes in luminal BC.

In TN BC, the presence of a NME surrounding the index mass, as illustrated in Figure 5, was significantly associated with non-pCR ($p$=0.032). No association was found between pCR and BES, tumor margins or shape.

In HER2-positive BC, no significant differences in the distribution of MRI features were observed between pCR and non pCR groups.

**MRI features and histological biomarkers**

As shown in Tables 4 and S5, spiculated margins and multifocality were significantly associated with low TILs levels in the whole population and in luminal BC. Similarly, a rim enhancement was significantly associated with low TILs levels in the whole population, in luminal BC, as well as in TN BC.

A lower background parenchymal enhancement was significantly associated with low TILs levels in the TN BC group.

Circumscribed or irregular margins and unifocality were significantly associated with high Ki67 levels in the whole population (Table 4). In the TN BC group, tumors with high Ki67 expression were more likely to be unifocal with no NME (Table S6).

**Discussion**

Our results show that non-spiculated margins, oval or round shape, smaller global MRI size, no associated non-mass enhancement and unifocality are associated with complete response after neoadjuvant chemotherapy on univariate analysis of the overall study population. Adjusting for known predictors of complete response such as BC subtype, T Stage, and tumor-infiltrating lymphocytes levels, multivariable analysis showed that non-spiculated margins and multifocality assessed by MRI remained independent predictors of complete response. We also showed that maximal MR size was an important variable using machine learning to predict pCR by random forest classifiers. The role of tumor size has been the subject of conflicting results in the literature [28, 29]. Some authors argue that molecular profile may be of greater importance for BC response than tumor size [30]. Still, tumor size is incorporated in decision trees for neoadjuvant chemotherapy from the ASCO guidelines, since numerous studies have shown lower rates of complete responses in larger tumors [31]. In a population including various subtypes of BC, Uematsu *et al.* reported that the MRI features associated with chemoresistant cancers were lesions without a mass effect, a larger tumor size and a high intra-tumoral signal intensity on T2-weighted images suggestive of intra-tumoral necrosis [8]. Kim *et al.* also showed that clinical T stage and lesion MRI size were associated with pCR on univariate analysis in a population of various BC subtypes [32]. We hypothesize that as we included tumor clinical stage in the multivariable analysis, the index tumor size may have not been selected for multivariable logistic regression because of its strong collinearity with the T Stage of the AJCC TNM staging system. Our results show that the maximal tumor size including the maximal extent of tumor foci is strongly associated with pCR, as it also indicates the participation of an associated non mass enhancement and/or multifocality, which are also associated with a poor response according to our results.

We showed that BES was not correlated with pCR but with index tumor size, not only in TN BC as reported by Harada *et al.* [18], but also in the population including all BC subtypes. BES is useful to diagnose occult inflammatory BC and to predict BC prognosis after NAC, as BES4 patients have significantly worse progression-free survival than BES1 patients [17]. Interestingly, we showed a significant association between BES4 class and lower TILS levels which are known to be associated with a poor prognosis [33]. TILs reflect host antitumor immunity and are a prognostic and predictive factor of BC outcome, particularly in TN and HER2 subtypes. Tumoral and peritumoral MRI morphologic and texture features have been studied to characterize the growing patterns and the tumor immune microenvironment of BC. Braman *et al.* evaluated radiomic features in the context of neoadjuvant chemotherapy for HER2-positive breast cancer and they showed that radiomics features could strongly predict pCR and that features from the peritumoral region were significantly associated with the TILs density [34]. A study performed on two breast cancer cohorts from The Cancer Genome Atlas (TCGA) project and the I-SPY 1 trial showed that TILs were significantly associated with specific enhancement patterns of tumor and surrounding parenchyma at DCE-MRI [35]. Another study reported TILs to be associated with peritumoral edema in HER2-positive BC [36].

Our study showed that TN BC surrounded by NME are associated with lower pCR rates and that NME is also correlated with lower Ki67 levels, which is known to be predictive of poor response to NAC. Shin *et al.* previously showed that the presence of NME on preoperative MRI was independently associated with worse locoregional recurrence-free survival in patients treated by breast-conserving surgery after NAC in a population comprising various BC subtypes, possibly due to underestimation of extent of microscopic tumor foci [37].

Eom *et al.* in a population of TN BC showed that oval or round shape and homogeneous enhancement on pretreatment MRI were associated with pCR on univariate analysis. On multivariate analysis, only homogeneous enhancement was significantly associated with pCR [9]. In their study, patients achieving near pCR with small residual invasive foci were included in the pCR group and the axilla was not included in the response assessment in contrast with the RCB used in the present study. The use of RCB to document pathologic response is recommended by the BIG-NABCG to standardize BC response evaluation, as it is an independent prognostic factor after NAC [38, 39]. Although we did not find any association between internal enhancement type and pCR, our results showed significantly lower TILs levels were associated with rim enhancement as previously reported in a population of all BC subtypes [40], and in TN BC as described by Ku *et al.* [41] but for the first time in the subgroup of luminal BC also.

We showed that spiculated margins and multifocality were significantly associated with non pCR in luminal BC and were associated with low TILs levels, which are predictive of non-response in luminal BC [42]. These MR patterns may help to better predict luminal BC response and better stratify these patients as their pCR rate remains low.

Because the threshold for KI67 has changed over time and is not uniformly defined, comparison with the literature regarding luminal BC response to NAC is challenging. Indeed, luminal BC patient populations enrolled in studies may vary significantly depending on the criterion used to define low KI67, and subsequently applied to distinguish low-proliferative luminal-A from high-proliferative luminal-B breast cancers, which have very different prognosis and response rates to NAC, the luminal-A subtype having the lowest response rates of all BC subtypes. Finally, our results showed a highly significant association in luminal BC between multifocality and low TILs level, but not in TN BC, as previously reported by Ku *et al.* [41], possibly due to lack of power, .

Quantification of computer extracted imaging features is expected to show better reproducibility and accuracy than qualitative BI-RADS characteristics assessed by the human eye. Radiomics studies results are very promising [43], yet these techniques remain difficult to implement in routine practice and are associated with low inter-center reproducibility because of a strong influence by the type of equipment and the acquisition parameters on their values [44]. Our results are in the range of previously published performances of MRI-based radiomics multivariable models for breast cancer response prediction as described in a literature review by Granzier *et al.* [45]. This review showed that the AUC results of multivariate analysis, ranged from 0.47 to 0.94 for validation cohorts.

This exploratory study has several limitations: it was a retrospective single-center cohort; analyses according to BC subtypes were limited by the small sample size; we included MRI acquired in other institutions. Although qualitative MRI features are less affected by variations in protocol parameters than computed features, these differences could have an impact on interpretation of the tumor and breast signal. Nevertheless, these aspects are representative of real-life MRI interpretations conditions. We found that Delong test was significant for 14 iterations, indicating that adding MRI parameters significantly improved the AUC in these iterations. However, in the other 86 iterations, the small size of the test set relative to the full data set limited our ability to confirm the null hypothesis. The Delong test has a potential to be overly conservative when analyzing improvements in prediction performance brought on by the inclusion of a new marker in exploratory studies [46]. Larger studies are thus needed to confirm our findings.

In conclusion, in a population including all breast cancer subtypes, spiculated margins and multifocality were independently associated with incomplete response as assessed by residual cancer burden, which is a consensual predictor of outcome after neoadjuvant chemotherapy. The total extent of tumor disease was also a strong predictor of response after NAC by machine learning classifiers. We confirmed the previously demonstrated association of BES with MR tumor size in TN BC, but also showed this association also applied to luminal BC and found that BES4 correlated with lower TILS levels. Our results demonstrated that spiculated margins are linked with a poor response and are associated with low TILS levels in the group of luminal BC, whose characteristics on MRI before neoadjuvant treatment have received limited research despite an urgent need to better stratify these patients who show poor response rates to NAC. Furthermore, a non-mass enhancement associated with TN BC was correlated with an incomplete response and low Ki67 index.

Further studies based on large data sets are thus needed to define the best use of these standardized MRI descriptors that are part of the radiologist clinical daily practice and highly transferable, to develop prediction models of complete response after neoadjuvant chemotherapy according to breast cancers subtypes, to improve the stratification of patients with locally advanced or aggressive breast cancer.

**TABLES LEGENDS**

Table 1. Breast MRI acquisition parameters used in our center

Table 2. Patients and tumor characteristics and their association with pathological complete response (pCR)

Table 3. Comparison of MRI features according to pathologic complete response (pCR) status by univariate logistic regression analysis

Table 4. Association between MRI features and biological predictors of pathologic complete response (pCR): TILs and KI67 levels

**FIGURES LEGENDS**

**Figure 1.** Study flowchart of participant selection

**Figure 2.** Tumor response according to residual cancer burden (RCB) classes presented by BC subtype for each MRI feature significantly associated with pathological complete response (pCR) on univariate analysis : margins **(A)**, shape **(B)**, associated non-mass enhancement **(C)**, and multifocality **(D)**.

**Figure 3.** Proportional values of pCR and patterns of MRI features are presented by breast cancer subtype. The heights of the boxes are proportional to the percentage of pCR and the width to the percentage of each pattern of MRI feature.

**Figure 4.** Luminal invasive not otherwise specified carcinoma in a 33-year-old patient, Ki67 index 30%, tumor-infiltrating lymphocytes (TILs) 5% before neoadjuvant chemotherapy (NAC), residual cancer burden (RCB) II (moderate residual disease) after NAC completion: **(A)** axial T2-weighted DIXON breast MRI scan shows a mass (arrowhead) of the posterior third of the breast, without peritumoral, prepectoral or subcutaneous edema, breast edema score (BES) 1; **(B)** axial contrast-enhanced T1-weighted breast MRI shows an irregular mass with spiculated margins and a rim enhancement, with an intra-tumoral coil seen as signal void (arrow); **(C)** sagittal contrast-enhanced T1-weighted breast MRI shows multifocal disease with additional masses (thick arrow) anterior to the index mass (thin arrow).

**Figure 5.** Triple negative invasive not otherwise specified breast cancer in a 37-year-old patient, Ki67 index 60%, tumor-infiltrating lymphocytes (TILs) 50%, residual cancer burden (RCB) II (moderate residual disease) after neoadjuvant chemotherapy: **(A)** axial T2-weighted DIXON breast MRI shows the tumor mass with an intratumoral high signal intensity (white curved arrow), and an intra-tumoral coil seen as signal void, BES (breast edema score) 1; Axial **(B)** and sagittal multiplanar reconstruction **(C)** T1-weighted contrast-enhanced breast MRI: an irregular mass is seen with irregular margins and a homogeneous enhancement (white thin arrows), with associated segmental nonmass enhancement oriented toward the nipple (white thick arrows).

**SUPPLEMENTAL MATERIAL**

**TABLES**

**Table S1.** Association between clinical and biological features and pathological complete response (pCR), according to the three breast cancer subtypes (Luminal, HER2-positive, and TN)

**Table S2.** Interreader agreement for the assessment of MRI features

**Table S3.** Results from multivariable logistic regression analysis to assess variables association with pCR

**Table S4.** Association between MRI features and pathological complete response (pCR), by breast cancer subtype (Luminal, HER2-positive, TN)

**Table S5.** Association between MRI features and tumor-infiltrating lymphocytes (TILs) levels (less or greater than 30%) by breast cancer subtype (Luminal, HER2-positive, TN)

**Table S6**. Association between MRI features and Ki67 levels by breast cancer subtype (Luminal, HER2-positive, TN)

**FIGURES**

**Figure S1.** Breast Edema Score illustrated by representative axial T2-weighted images. No edema is visible surrounding the tumor depicted by a curved arrow, BES1 (A). Image shows a peritumoral edema appearing as a high signal intensity surrounding an irregular mass (straight arrows), corresponding to BES 2 category (B). Prepectoral edema (BES 3) is seen as a high signal intensity is seen between the tumor and the pectoralis major muscle (thick arrow) (C). Subcutaneous edema (BES4) appears as a high-signal-intensity in the subcutaneous area (arrowheads) (D).

**Figure S2.** Index Lesion MRI size distribution according to Breast Edema Score (BES) in the whole population **(A)**, in Luminal **(B)**, HER2-positive **(C)**, and TN BC **(D)**. Significant differences in MR size were shown according to BES in the whole population, as well as in luminal and TN BC. In the whole population, pairwise analyses with correction for multiple comparisons showed significant differences between BES 1, and BES 4 (*p*=0.001), BES 3 (*p*=0.002). Significant differences between BES 1 and BES 4 were also shown in Luminal BC (*p*=0.024).

**Figure S3.** TILs expression levels by the absence (BES 1/2/3) and presence (BES 4) of a subcutaneous edema on T2-weighted MRI

**Figure S4.** Forest plot displaying the Odds Ratios derived from multivariable logistic regression analysis.

**Figure S5.** Graph derived from multiple component analysis. Both subjects (dots) and variables (arrows) appear on the plot. Ellipses delineate the pCR (light green) and non pCR groups (dark green). The graph shows how low KI67 and low TILs levels are correlated and mostly associated with spiculated margins as the arrows from these variables are in close proximity. On the other hand, unifocality and the absence of an associated NME are closely associated with TN subtype and high TILs levels.

Table 1

Table 1. Breast MRI acquisition parameters in our center

| Coil | Coil 1[*], N=14 | | Coil 2[**], N=55 | | Coil 3[***], N=22 | |
|---|---|---|---|---|---|---|
| | T2-weighted sequence | T1-weighted DCE sequence | T2-weighted sequence | T1-weighted DCE sequence | T2-weighted sequence | T1-weighted DCE sequence |
| TR (ms) | 3310 | 5.2 | 6400 | 5.2 | 5544 | 6.81 |
| TE (ms) | 88 | 2.4 | 88 | 2.4 | 90 | 3.3 |
| Slice thickness (mm) | 3.5 | 0.9 | 3 | 0.9 | 3 | 1 |
| Slice interval (mm) | 4.2 | 0.9 | 3.6 | 0.9 | 3.3 | 1 |
| Pixel spacing (mm) | 0.7x0.7 | 0.91x0.91 | 0.7x0.7 | 0.91x0.91 | 0.68x0.68 | 0.82x0.82 |
| Pixel bandwidth (Hz/pix) | 315 | 355 | 375 | 355 | 558 | 434 |
| Flip angle (°) | 150 | 10 | 180 | 10 | 160 | 15 |
| Field of View (mm$^2$) | 360x360 | 380x342 | 360x360 | 380x380 | 350x350 | 420x420 |
| Matrix | 512x435 | 416x312 | 512x435 | 416x312 | 416x416 | 416x416 |

[*]Coil 1: 18-channel coil, 1.5 T magnet, MAGNETOM Aera (Siemens)
[**]Coil 2: 16-channel coil (Sentinelle; Philips Medical Systems), MAGNETOM Aera (Siemens)
[***]Coil 3: 8-channel coil, 1.5 T magnet, Optima MR450w (GE)
DCE: dynamic contrast-enhanced, TE: time to echo, TR: repetition time

Table 2

Table 2. Patients and tumor characteristics and their association with pathological complete response (pCR)

| Features | Overall, N = 129[1] | non pCR, N = 70[1] | pCR, N = 59[1] | p-value[2] |
|---|---|---|---|---|
| Age | 47 (39, 55) | 48 (40, 56) | 46 (38, 54) | 0.380 |
| Body Mass Index | 23 (21, 26) | 23 (21, 26) | 23 (21, 26) | 0.928 |
| Menopausal Status | | | | 0.214 |
|     Postmenopausal | 49 | 30 (61%) | 19 (39%) | |
|     Premenopausal | 80 | 40 (50%) | 40 (50%) | |
| T Stage | | | | **0.008** |
|     0/I/II | 111 | 55 (50%) | 56 (50%) | |
|     III/IV | 18 | 15 (83%) | 3 (17%) | |
| N Stage | | | | 0.275 |
|     0 | 72 | 36 (50%) | 36 (50%) | |
|     I/II/III | 57 | 34 (60%) | 23 (40%) | |
| M Stage | | | | >0.999 |
|     0 | 127 | 69 (54%) | 58 (46%) | |
|     I | 2 | 1 (50%) | 1 (50%) | |
| Tumor type | | | | 0.853 |
|     Ductal NOS | 125 | 68 (54%) | 57 (46%) | |
|     Lobular | 1 | 1 (100%) | 0 (0%) | |
|     Mixt | 1 | 0 (0%) | 1 (100%) | |
|     Other | 2 | 1 (50%) | 1 (50%) | |
| Breast Cancer Subtype | | | | **<0.001** |
|     Luminal | 37 | 30 (81%) | 7 (19%) | |
|     HER2-positive | 37 | 15 (41%) | 22 (59%) | |
|     TNBC | 55 | 25 (45%) | 30 (55%) | |
| Grade | | | | 0.532 |
|     2 | 43 | 25 (58%) | 18 (42%) | |
|     3 | 86 | 45 (52%) | 41 (48%) | |
| Ki 67 % | 50 (30.70) | 40 (30, 70) | 60 (40, 80) | **0.005** |
| TILs % | 20 (10,40) | 15 (5, 38) | 30 (10, 50) | **0.016** |

[1] n (%) or Median (inter quartile range) according to the features
[2] Fisher's exact test, Pearson's Chi-squared test or Wilcoxon rank sum test according to the features
NOS: Not Otherwise Specified

Table 3

Table 3. Comparison of MRI features according to pathologic complete response (pCR) status by univariate logistic regression analyses

| Characteristic | Overall, N = 129[1] | non pCR, N = 70[1] | pCR, N = 59[1] | OR* | 95% CI* | p-value |
|---|---|---|---|---|---|---|
| Margins | | | | | | **0.018** |
|     Circumscribed / Irregular | 73 | 33 (45%) | 40 (55%) | — | — | |
|     Spiculated | 56 | 37 (66%) | 19 (34%) | 0.42 | 0.20, 0.86 | |
| Shape | | | | | | **0.047** |
|     Irregular | 96 | 57 (59%) | 39 (41%) | — | — | |
|     Oval / Round | 33 | 13 (39%) | 20 (61%) | 2.25 | 1.01, 5.14 | |
| Intratumoral high signal intensity on T2 | | | | | | 0.648 |
|     Present | 42 | 24 (57%) | 18 (43%) | — | — | |
|     Absent | 87 | 46 (53%) | 41 (47%) | 1.19 | 0.57, 2.52 | |
| Peritumoral Edema | | | | | | 0.925 |
|     Present | 88 | 48 (55%) | 40 (45%) | — | — | |
|     Absent | 41 | 22 (54%) | 19 (46%) | 1.04 | 0.49, 2.18 | |
| Prepectoral Edema | | | | | | 0.484 |
|     Absent | 85 | 48 (56%) | 37 (44%) | — | — | |
|     Present | 44 | 22 (50%) | 22 (50%) | 1.30 | 0.62, 2.70 | |
| Subcutaneous Edema | | | | | | 0.075 |
|     Present | 16 | 12 (75%) | 4 (25%) | — | — | |
|     Absent | 113 | 58 (51%) | 55 (49%) | 2.84 | 0.93, 10.7 | |
| BES | | | | | | 0.214 |
|     1 | 36 | 19 (53%) | 17 (47%) | — | — | |
|     2 | 44 | 25 (57%) | 19 (43%) | 0.85 | 0.35, 2.06 | |
|     3 | 33 | 14 (42%) | 19 (58%) | 1.52 | 0.59, 3.98 | |
|     4 | 16 | 12 (75%) | 4 (25%) | 0.37 | 0.09, 1.30 | |
| Multifocality | | | | | | **0.026** |
|     Present | 29 | 21 (72%) | 8 (28%) | — | — | |
|     Absent | 100 | 49 (49%) | 51 (51%) | 2.73 | 1.14, 7.10 | |
| Background Parenchymal Enhancement | | | | | | 0.730 |
|     Minimal/Mild | 101 | 54 (53%) | 47 (47%) | — | — | |
|     Moderate/Marked | 28 | 16 (57%) | 12 (43%) | 0.86 | 0.36, 2.00 | |
| Associated non-mass Enhancement | | | | | | **0.024** |
|     Absent | 105 | 52 (50%) | 53 (50%) | — | — | |
|     Present | 24 | 18 (75%) | 6 (25%) | 0.33 | 0.11, 0.85 | |
| Internal Enhancement Type | | | | | | 0.232 |
|     Homogeneous / Heterogeneous | 94 | 48 (51%) | 46 (49%) | — | — | |
|     Rim Enhancement | 35 | 22 (63%) | 13 (37%) | 0.62 | 0.27, 1.35 | |
| Delayed Phase Enhancement | | | | | | 0.775 |
|     Persistent / Plateau | 41 | 23 (56%) | 18 (44%) | — | — | |
|     Wash-out | 88 | 47 (53%) | 41 (47%) | 1.11 | 0.53, 2.37 | |
| Breast Composition | | | | | | 0.609 |
|     A-B | 60 | 34 (57%) | 26 (43%) | — | — | |
|     C-D | 69 | 36 (52%) | 33 (48%) | 1.20 | 0.60, 2.41 | |
| Index Lesion MR Size | 26 (22, 35) | 26 (23, 37) | 25 (21, 34) | 0.99 | 0.95, 1.02 | 0.372 |
| Maximal MR Size | 30 (23, 40) | 32 (24, 42) | 27 (22, 36) | 0.98 | 0.96, 1.00 | **0.031** |

*OR = Odds Ratio, CI = Confidence Interval

[1] n (%); Median (IQR)

BES: breast edema score, pCR: pathologic complete response, TILs: tumor-infiltrating lymphocytes

Table 4

Table 4 Association between MRI features and biological predictors of pathologic complete response (pCR): TILs and KI67 levels

|  | TILs | | | KI67 | | |
|---|---|---|---|---|---|---|
| **Characteristic** | TILs < 30% N = 72[1] | TILs ≥ 30% N = 57[1] | *p* value[2] | KI67 ≤ 25% N = 16[1] | KI67 > 25% N = 113[1] | *p* value[2] |
| Margins | | | **0.006** | | | **0.006** |
|     Circumscribed / Irregular | 33 (45%) | 40 (55%) | | 4 (5.5%) | 69 (95%) | |
|     Spiculated | 39 (70%) | 17 (30%) | | 12 (21%) | 44 (79%) | |
| Shape | | | 0.813 | | | >0.999 |
|     Irregular | 53 (55%) | 43 (45%) | | 12 (12%) | 84 (88%) | |
|     Oval / Round | 19 (58%) | 14 (42%) | | 4 (12%) | 29 (88%) | |
| Intratumoral high signal intensity on T2 | | | 0.085 | | | 0.491 |
|     Present | 28 (67%) | 14 (33%) | | 4 (9.5%) | 38 (90%) | |
|     Absent | 44 (51%) | 43 (49%) | | 12 (14%) | 75 (86%) | |
| BES | | | **0.050** | | | 0.461 |
|     1 | 19 (53%) | 17 (47%) | | 3 (8.3%) | 33 (92%) | |
|     2 | 21 (48%) | 23 (52%) | | 5 (11%) | 39 (89%) | |
|     3 | 18 (55%) | 15 (45%) | | 5 (15%) | 28 (85%) | |
|     4 | 14 (88%) | 2 (12%) | | 3 (19%) | 13 (81%) | |
| Multifocality | | | **0.041** | | | **0.050** |
|     Present | 21 (72%) | 8 (28%) | | 7 (24%) | 22 (76%) | |
|     Absent | 51 (51%) | 49 (49%) | | 9 (9.0%) | 91 (91%) | |
| Background Parenchymal Enhancement | | | 0.258 | | | 0.749 |
|     Minimal/Mild | 59 (58%) | 42 (42%) | | 12 (12%) | 89 (88%) | |
|     Moderate/Marked | 13 (46%) | 15 (54%) | | 4 (14%) | 24 (86%) | |
| Associated non-mass Enhancement | | | 0.101 | | | 0.177 |
|     Absent | 55 (52%) | 50 (48%) | | 11 (10%) | 94 (90%) | |
|     Present | 17 (71%) | 7 (29%) | | 5 (21%) | 19 (79%) | |
| Internal Enhancement Type | | | **0.003** | | | 0.135 |
|     Homogeneous / Heterogeneous | 45 (48%) | 49 (52%) | | 9 (9.6%) | 85 (90%) | |
|     Rim Enhancement | 27 (77%) | 8 (23%) | | 7 (20%) | 28 (80%) | |
| Delayed Phase Enhancement | | | 0.235 | | | 0.534 |
|     Persistent / Plateau | 26 (63%) | 15 (37%) | | 4 (9.8%) | 37 (90%) | |
|     Wash-out | 46 (52%) | 42 (48%) | | 12 (14%) | 76 (86%) | |
| Breast Composition | | | 0.591 | | | 0.765 |
|     A-B | 35 (58%) | 25 (42%) | | 8 (13%) | 52 (87%) | |
|     C-D | 37 (54%) | 32 (46%) | | 8 (12%) | 61 (88%) | |
| Index Lesion MR Size | 25 (24, 33) | 25 (21, 33) | 0.442 | 24 (22, 36) | 27 (22, 35) | 0.991 |
| Maximal MR Size | 33 (24, 41) | 30 (24, 36) | 0.290 | 35 (24, 48) | 30 (23, 40) | 0.520 |

[1] n (%) or Median (inter quartile range) according to the features

[2] Fisher's exact test, Pearson's Chi-squared test or Wilcoxon rank sum

BES: breast edema score, pCR: pathologic complete response, TILs: tumor-infiltrating lymphocytes

Figure 1

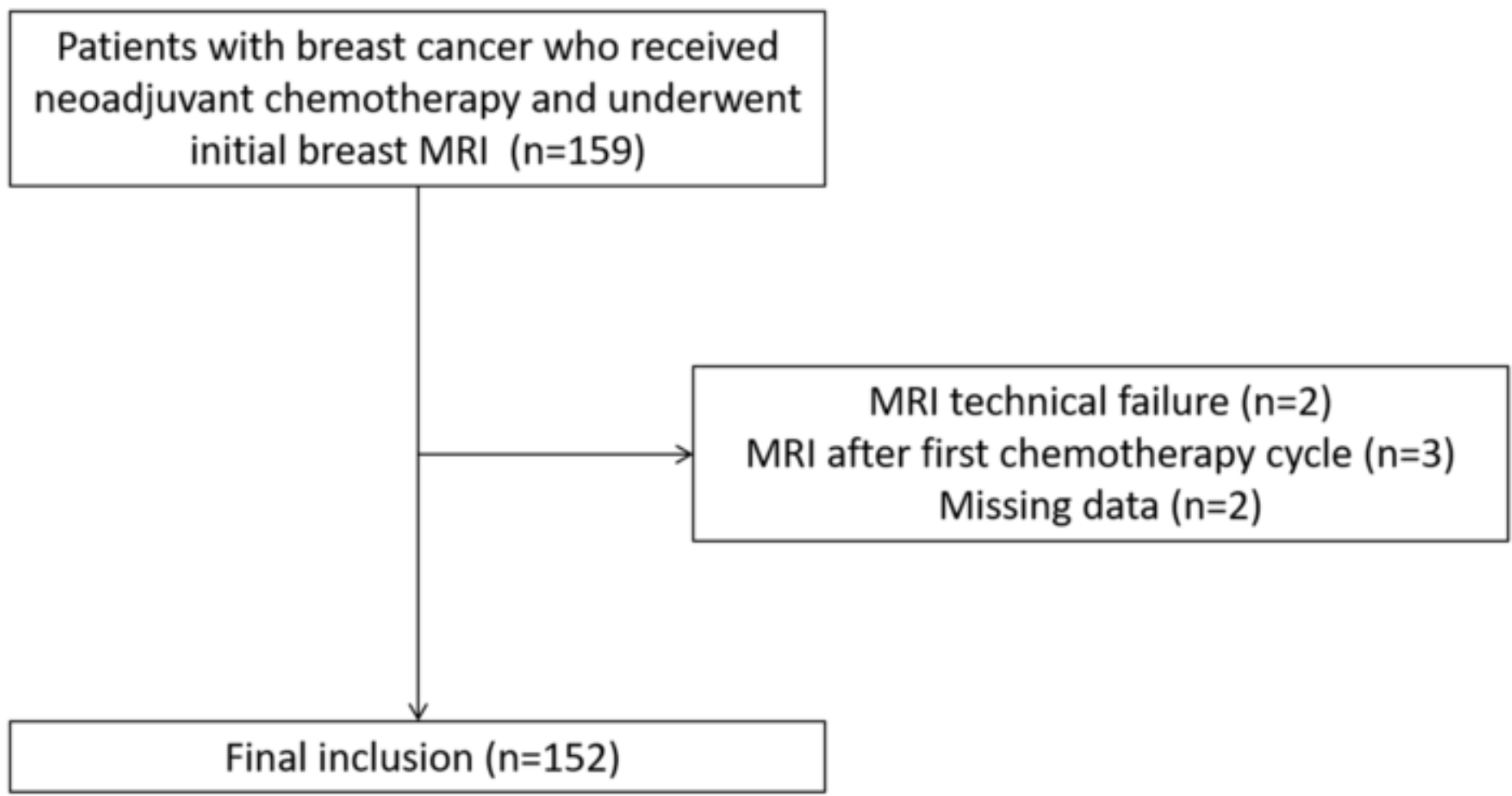

Figure 2

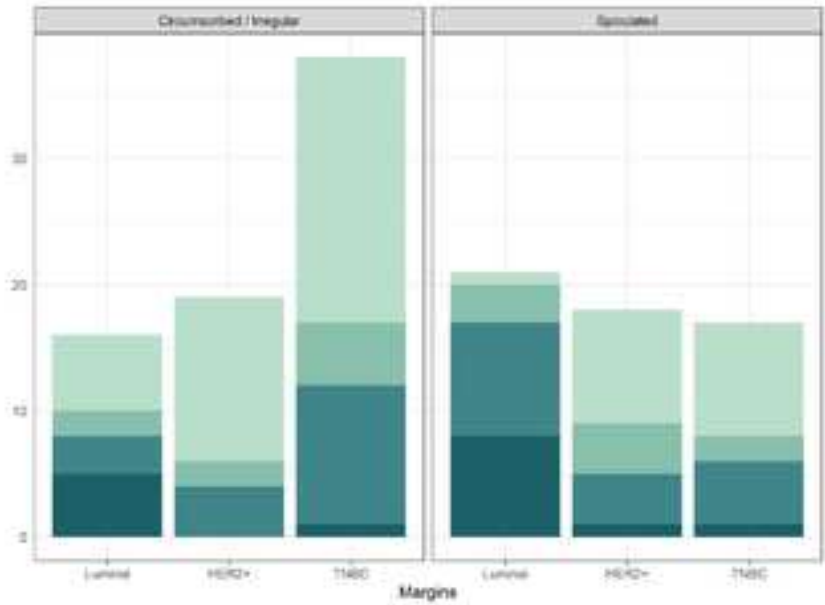
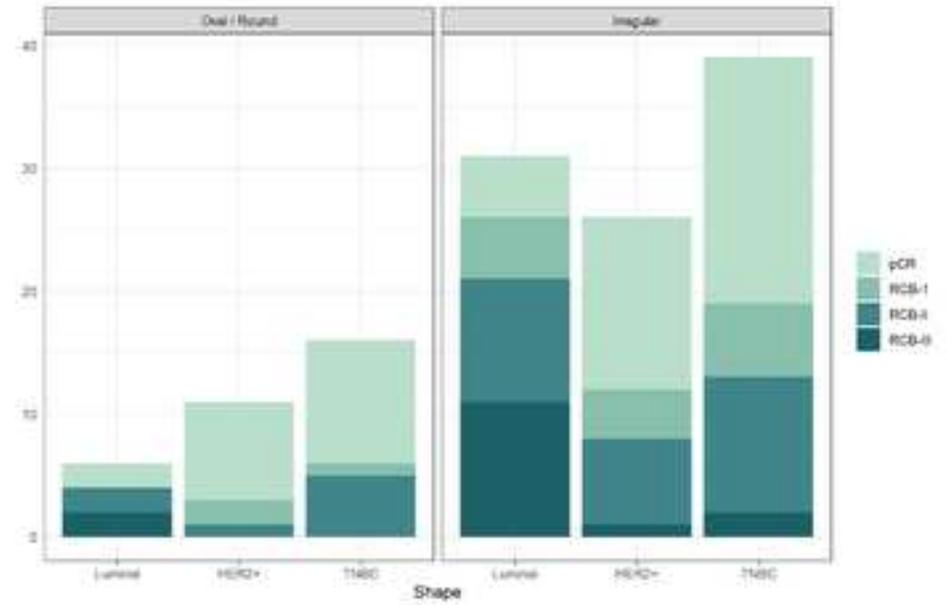
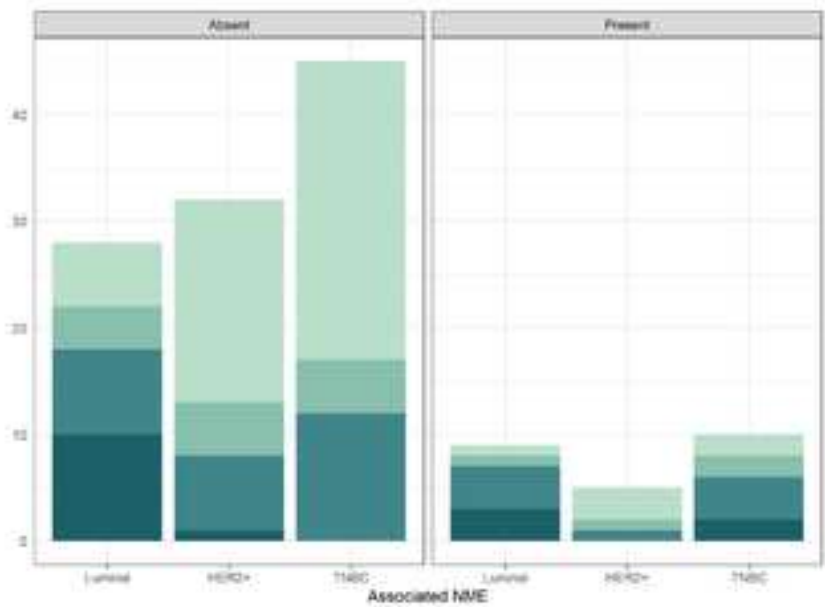
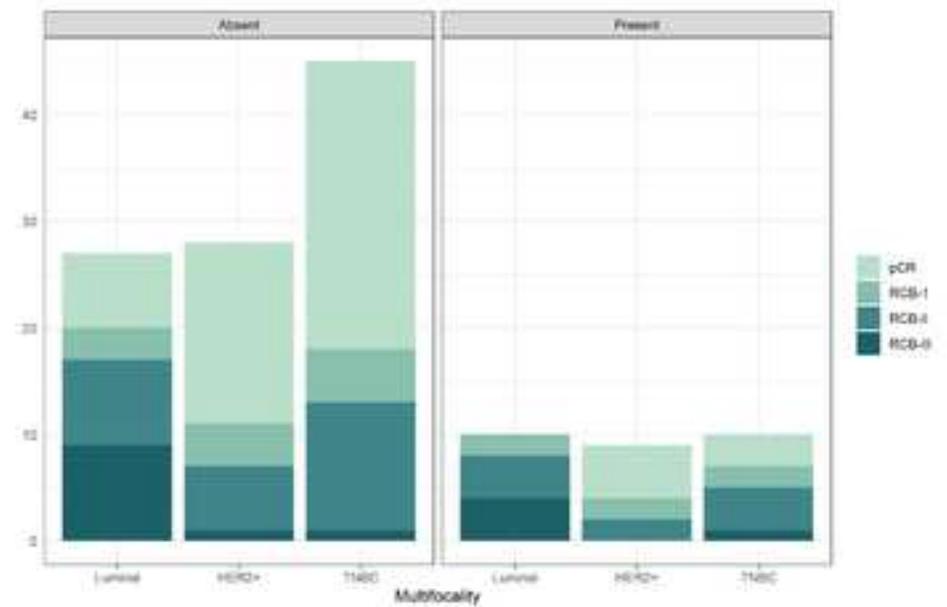

Figure 3

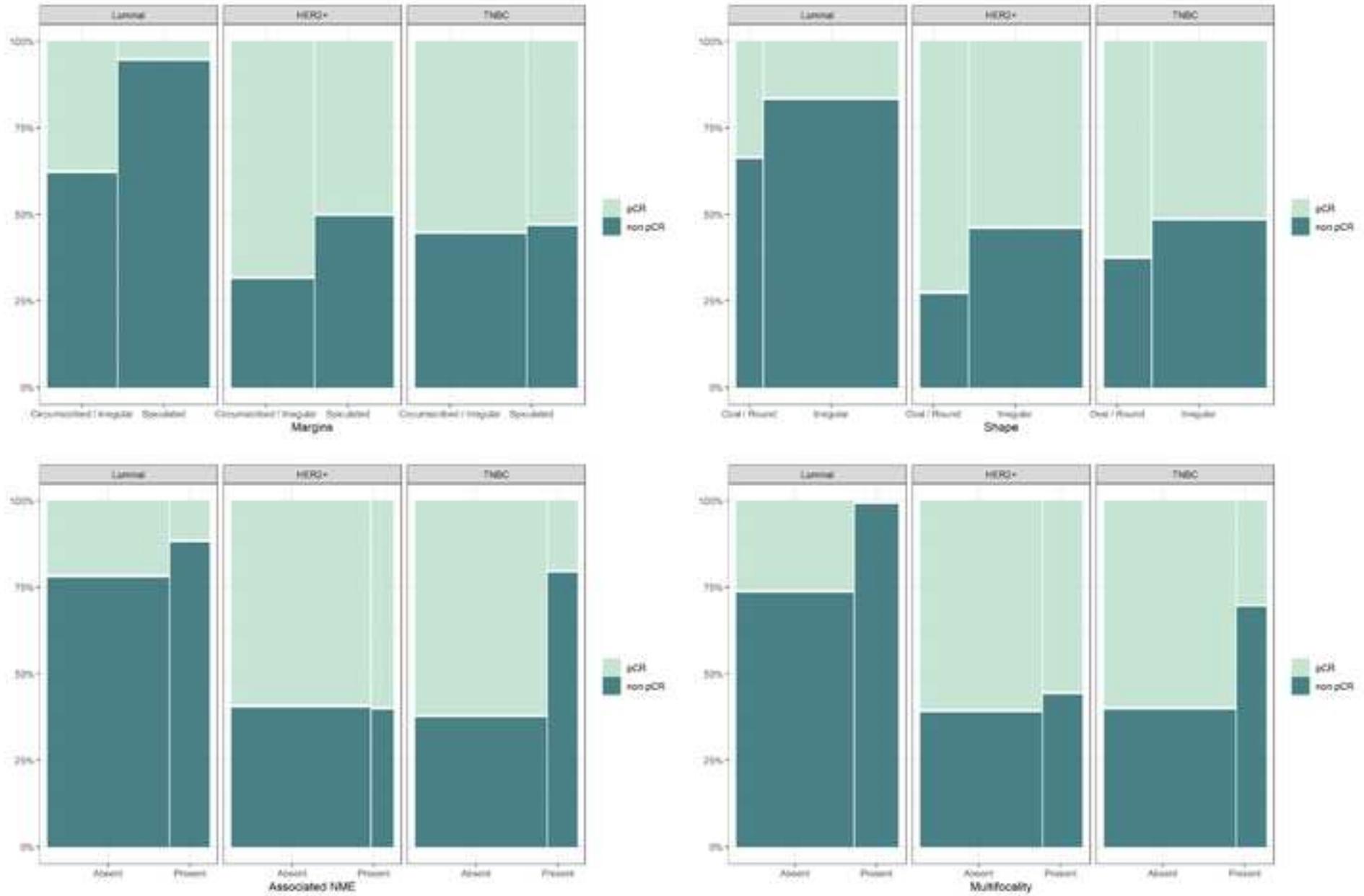

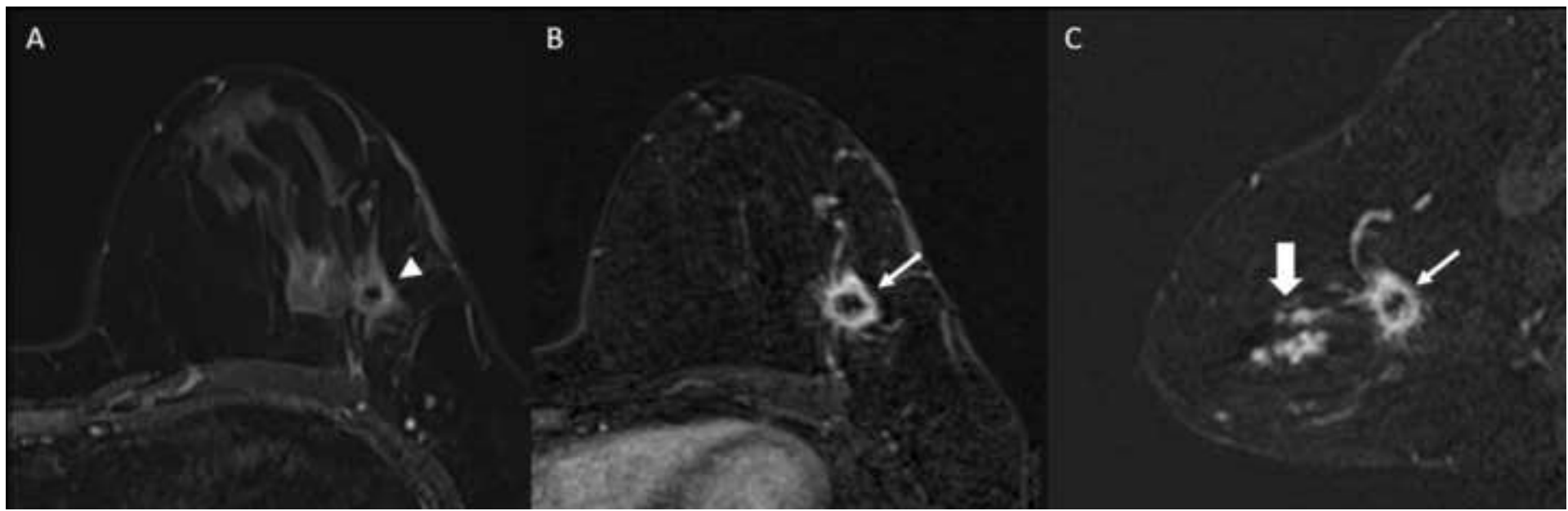

Figure 5

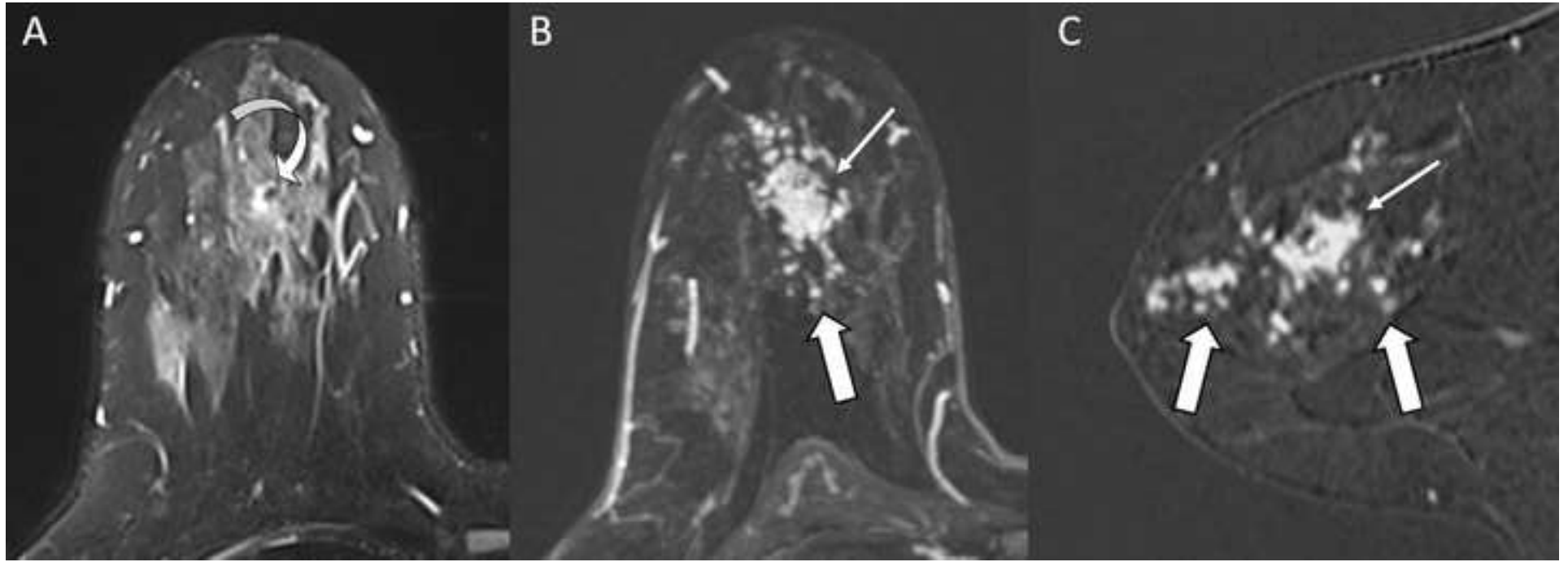

**Table S1.** Association between clinical and biological features and pathological complete response (pCR), according to the three breast cancer subtypes (Luminal, HER2-positive, and TN)

| | Luminal | | | | HER2 | | | | TN | | | |
|---|---|---|---|---|---|---|---|---|---|---|---|---|
| Characteristic | Overall N = 37[1] | non pCR N = 30[1] | pCR N = 7[1] | p-value[2] | Overall N = 37[1] | non pCR N = 15[1] | pCR N = 22[1] | p-value[2] | Overall N = 55[1] | non pCR N = 25[1] | pCR N = 30[1] | p-value[2] |
| Age | 47 (39, 56) | 48 (42, 56) | 40 (37, 42) | 0.116 | 48 (43, 58) | 48 (43, 58) | 48 (40, 55) | 0.951 | 47 (38, 54) | 48 (39, 52) | 46 (37, 54) | 0.872 |
| BMI | 24.6 (22.1, 28.4) | 25.0 (22.1, 27.8) | 24.2 (22.6, 26.8) | 0.831 | 23.2 (20.7, 26.4) | 22.2 (20.4, 25.5) | 23.9 (21.1, 26.3) | 0.293 | 23.0 (21.6, 24.8) | 22.9 (21.5, 24.6) | 23.3 (21.6, 24.8) | 0.960 |
| Menopausal Status | | | | 0.113 | | | | 0.531 | | | | 0.916 |
| Postmenopausal | 16 | 15 (94%) | 1 (6.2%) | | 15 | 7 (47%) | 8 (53%) | | 18 | 8 (44%) | 10 (56%) | |
| Premenopausal | 21 | 15 (71%) | 6 (29%) | | 22 | 8 (36%) | 14 (64%) | | 37 | 17 (46%) | 20 (54%) | |
| T Stage | | | | 0.308 | | | | 0.377 | | | | 0.165 |
| 0/I/II | 29 | 22 (76%) | 7 (24%) | | 32 | 12 (38%) | 20 (62%) | | 50 | 21 (42%) | 29 (58%) | |
| III/IV | 8 | 8 (100%) | 0 (0%) | | 5 | 3 (60%) | 2 (40%) | | 5 | 4 (80%) | 1 (20%) | |
| N Stage | | | | **0.008** | | | | 0.315 | | | | 0.765 |
| 0 | 19 | 12 (63%) | 7 (37%) | | 21 | 10 (48%) | 11 (52%) | | 32 | 14 (44%) | 18 (56%) | |
| I/II/III | 18 | 18 (100%) | 0 (0%) | | 16 | 5 (31%) | 11 (69%) | | 23 | 11 (48%) | 12 (52%) | |
| M Stage | | | | >0.999 | | | | >0.999 | | | | >0.999 |
| 0 | 36 | 29 (81%) | 7 (19%) | | 36 | 15 (42%) | 21 (58%) | | 55 | 25 (45%) | 30 (55%) | |
| I | 1 | 1 (100%) | 0 (0%) | | 1 | 0 (0%) | 1 (100%) | | 0 | 0 (NA%) | 0 (NA%) | |
| Tumor type | | | | >0.999 | | | | >0.999 | | | | >0.999 |
| Ductal NOS | 36 | 29 (81%) | 7 (19%) | | 34 | 14 (41%) | 20 (59%) | | 55 | 25 (45%) | 30 (55%) | |
| Lobular | 1 | 1 (100%) | 0 (0%) | | 0 | 0 (NA%) | 0 (NA%) | | 0 | 0 (NA%) | 0 (NA%) | |
| Mixt | 0 | 0 (NA%) | 0 (NA%) | | 1 | 0 (0%) | 1 (100%) | | 0 | 0 (NA%) | 0 (NA%) | |
| Other | 0 | 0 (NA%) | 0 (NA%) | | 2 | 1 (50%) | 1 (50%) | | 0 | 0 (NA%) | 0 (NA%) | |
| Grade | | | | 0.389 | | | | 0.457 | | | | 0.176 |
| 2 | 12 | 11 (92%) | 1 (8.3%) | | 20 | 7 (35%) | 13 (65%) | | 11 | 7 (64%) | 4 (36%) | |
| 3 | 25 | 19 (76%) | 6 (24%) | | 17 | 8 (47%) | 9 (53%) | | 44 | 18 (41%) | 26 (59%) | |

Eur Radiol (2023) Malhaire C, Selhane F, Saint-Martin MJ et al

| | | | | | | | | | | | | |
|---|---|---|---|---|---|---|---|---|---|---|---|---|
| Ki 67 | 40 (30, 60) | 38 (30, 58) | 50 (38, 70) | 0.166 | 40 (30, 60) | 30 (28, 55) | 40 (30, 68) | 0.575 | 70 (50, 80) | 60 (40, 80) | 70 (60, 80) | **0.077** |
| TILs % | 20 (5, 40) | 15 (5, 30) | 30 (30, 40) | 0.107 | 15 (10, 30) | 20 (10, 40) | 15 (6, 20) | 0.511 | 30 (10, 55) | 15 (5, 30) | 40 (26, 60) | **0.006** |

1 Median (IQR); n (%)
2 Wilcoxon rank sum test; Pearson's Chi-squared test; Fisher's exact test

Table S2. A) Inter-reader agreement

| MRI Feature | Agreement (%) | Kappa value |
|---|---|---|
| Intratumoral high signal intensity on T2 | 92.9 | 0.84 |
| Associated NME | 98.4 | 0.95 |
| Pre-pectoral Edema | 95.3 | 0.89 |
| Subcutaneous Edema | 97.6 | 0.90 |
| Peritumoral Edema | 96.9 | 0.93 |
| Internal Enhancement* | 85.8 | 0.81 |
| Shape* | 88.2 | 0.81 |
| Margins* | 91.3 | 0.86 |

CI: confidence interval, MRI: magnetic resonance imaging, NME: nonmass enhancement

A κ value of 0 represents no agreement; a κ value of 0.2 or less, slight agreement; a κ value of 0.21-0.40, fair agreement; a κ value of 0.41-0.60, moderate agreement; a κ value of 0.61-0.80, substantial agreement, and a κ value of 0.81-0.99, excellent agreement.
*linear-weighted K was used for ordinal variables

Eur Radiol (2023) Malhaire C, Selhane F, Saint-Martin MJ et al

Table S2. B-) Raw agreement percentages for ordinal MRI features

| Margins | Circumscribed | Irregular | Spiculated |
|---|---|---|---|
| Circumscribed | 2 | 1 | 0 |
| Irregular | 0 | 51 | 2 |
| Spiculated | 0 | 6 | 38 |

| Shape | Oval | Round | Irregular |
|---|---|---|---|
| Oval | 10 | 1 | 1 |
| Round | 5 | 8 | 1 |
| Irregular | 1 | 2 | 70 |

| Internal Enhancement | Homogeneous | Heterogeneous | Rim enhancement |
|---|---|---|---|
| Homogeneous | 21 | 5 | 2 |
| Heterogeneous | 2 | 38 | 4 |
| Rim enhancement | 0 | 1 | 27 |

Table S3. Results from multivariate logistic regression analysis to assess variables association with pCR

| Characteristic | OR[1] | 95% CI[1] | p-value |
|---|---|---|---|
| **Breast Cancer Subtype** | | | **<0.001** |
| Luminal | — | — | |
| HER2+ | 9.95 | 3.06, 37.2 | |
| TNBC | 4.44 | 1.53, 14.2 | |
| **TILs** | | | **0.031** |
| High | — | — | |
| Low | 0.39 | 0.16, 0.92 | |
| **T Stage** | | | **0.002** |
| 0/I/II | — | — | |
| III/IV | 0.12 | 0.02, 0.48 | |
| **Multifocality** | | | **0.038** |
| Present | — | — | |
| Absent | 3.03 | 1.06, 9.40 | |
| **Margins** | | | **0.045** |
| Circumscribed / Irregular | — | — | |
| Spiculated | 0.42 | 0.17, 0.98 | |
| BMI | 1.08 | 0.97, 1.21 | 0.15 |

[1]OR = Odds Ratio, CI = Confidence Interval

Table S4. Association between MRI features and pathological complete response (pCR), by breast cancer subtype (Luminal, HER2-positive, TN)

| Characteristic | Luminal | | | HER2 | | | TN | | |
|---|---|---|---|---|---|---|---|---|---|
| | non pCR, N = 30[1] | pCR, N = 7[1] | p value[2] | non pCR, N = 15[1] | pCR, N = 22[1] | p value[2] | non pCR, N = 25[1] | pCR, N = 30[1] | p value[2] |
| Margins | | | **0.029** | | | 0.254 | | | 0.873 |
|     Circumscribed / Irregular | 10 (62%) | 6 (38%) | | 6 (32%) | 13 (68%) | | 17 (45%) | 21 (55%) | |
|     Spiculated | 20 (95%) | 1 (4.8%) | | 9 (50%) | 9 (50%) | | 8 (47%) | 9 (53%) | |
| Shape | | | 0.315 | | | 0.466 | | | 0.448 |
|     Irregular | 26 (84%) | 5 (16%) | | 12 (46%) | 14 (54%) | | 19 (49%) | 20 (51%) | |
|     Oval / Round | 4 (67%) | 2 (33%) | | 3 (27%) | 8 (73%) | | 6 (38%) | 10 (62%) | |
| Intratumoral high signal intensity on T2 | | | >0.999 | | | 0.724 | | | 0.456 |
|     Present | 11 (85%) | 2 (15%) | | 4 (33%) | 8 (67%) | | 9 (53%) | 8 (47%) | |
|     Absent | 19 (79%) | 5 (21%) | | 11 (44%) | 14 (56%) | | 16 (42%) | 22 (58%) | |
| Peritumoral Edema | | | 0.437 | | | 0.641 | | | >0.999 |
|     Present | 18 (86%) | 3 (14%) | | 10 (43%) | 13 (57%) | | 20 (45%) | 24 (55%) | |
|     Absent | 12 (75%) | 4 (25%) | | 5 (36%) | 9 (64%) | | 5 (45%) | 6 (55%) | |
| Prepectoral Edema | | | >0.999 | | | 0.466 | | | >0.999 |
|     Absent | 21 (81%) | 5 (19%) | | 12 (46%) | 14 (54%) | | 15 (45%) | 18 (55%) | |
|     Present | 9 (82%) | 2 (18%) | | 3 (27%) | 8 (73%) | | 10 (45%) | 12 (55%) | |
| Subcutaneous Edema | | | 0.560 | | | >0.999 | | | 0.226 |
|     Present | 5 (100%) | 0 (0%) | | 2 (50%) | 2 (50%) | | 5 (71%) | 2 (29%) | |
|     Absent | 25 (78%) | 7 (22%) | | 13 (39%) | 20 (61%) | | 20 (42%) | 28 (58%) | |
| BES | | | 0.461 | | | 0.817 | | | 0.469 |
|     1 | 10 (71%) | 4 (29%) | | 4 (33%) | 8 (67%) | | 5 (50%) | 5 (50%) | |
|     2 | 10 (91%) | 1 (9.1%) | | 6 (50%) | 6 (50%) | | 9 (43%) | 12 (57%) | |
|     3 | 5 (71%) | 2 (29%) | | 3 (33%) | 6 (67%) | | 6 (35%) | 11 (65%) | |
|     4 | 5 (100%) | 0 (0%) | | 2 (50%) | 2 (50%) | | 5 (71%) | 2 (29%) | |
| Multifocality | | | 0.155 | | | >0.999 | | | 0.158 |
|     Present | 10 (100%) | 0 (0%) | | 4 (44%) | 5 (56%) | | 7 (70%) | 3 (30%) | |

Eur Radiol (2023) Malhaire C, Selhane F, Saint-Martin MJ et al

| | | | | | | | | | |
|---|---|---|---|---|---|---|---|---|---|
| | Absent | 20 (74%) | 7 (26%) | | 11 (39%) | 17 (61%) | | 18 (40%) | 27 (60%) | |
| Background Parenchymal Enhancement | | | | >0.999 | | | 0.677 | | | 0.498 |
| | Minimal/Mild | 22 (81%) | 5 (19%) | | 13 (43%) | 17 (57%) | | 19 (43%) | 25 (57%) | |
| | Moderate/Marked | 8 (80%) | 2 (20%) | | 2 (29%) | 5 (71%) | | 6 (55%) | 5 (45%) | |
| Associated non-mass Enhancement | | | | 0.656 | | | >0.999 | | | **0.032** |
| | Absent | 22 (79%) | 6 (21%) | | 13 (41%) | 19 (59%) | | 17 (38%) | 28 (62%) | |
| | Present | 8 (89%) | 1 (11%) | | 2 (40%) | 3 (60%) | | 8 (80%) | 2 (20%) | |
| Internal Enhancement Type | | | | 0.155 | | | >0.999 | | | 0.456 |
| | Homogeneous / Heterogeneous | 20 (74%) | 7 (26%) | | 12 (41%) | 17 (59%) | | 16 (42%) | 22 (58%) | |
| | Rim Enhancement | 10 (100%) | 0 (0%) | | 3 (38%) | 5 (62%) | | 9 (53%) | 8 (47%) | |
| Delayed Phase Enhancement | | | | >0.999 | | | 0.956 | | | 0.692 |
| | Persistent / Plateau | 10 (83%) | 2 (17%) | | 6 (40%) | 9 (60%) | | 7 (50%) | 7 (50%) | |
| | Wash-out | 20 (80%) | 5 (20%) | | 9 (41%) | 13 (59%) | | 18 (44%) | 23 (56%) | |
| Breast Composition | | | | 0.113 | | | 0.306 | | | 0.349 |
| | A-B | 15 (94%) | 1 (6.2%) | | 8 (50%) | 8 (50%) | | 11 (39%) | 17 (61%) | |
| | C-D | 15 (71%) | 6 (29%) | | 7 (33%) | 14 (67%) | | 14 (52%) | 13 (48%) | |
| Maximal MR Size | | 32 (25, 41) | 22 (21, 30) | **0.030** | 30 (24, 41) | 29 (23, 36) | 0.757 | 33 (26, 46) | 28 (21, 36) | 0.190 |
| Index Lesion MR Size | | 26 (24, 36) | 22 (18, 23) | **0.022** | 24 (20, 38) | 28 (22, 35) | 0.598 | 31 (23, 37) | 27 (21, 34) | 0.624 |

[1] n (%); Median (IQR)

[2] Fisher's exact test; Wilcoxon rank sum test

Eur Radiol (2023) Malhaire C, Selhane F, Saint-Martin MJ et al

Table S5 Association between MRI features and tumor-infiltrating lymphocytes (TILs) levels < or ≥ 30% by breast cancer subtype (Luminal, HER2-positive)

| | Luminal | | | HER2 | | | TN | | |
|---|---|---|---|---|---|---|---|---|---|
| **Characteristic** | < 30% TILs N = 21[1] | ≥ 30% TILs N = 16[1] | *p* value[2] | < 30% TILs N = 26[1] | ≥ 30% TILs N = 11[1] | *p* value[2] | < 30% TILs N = 25[1] | ≥ 30% TILs N = 30[1] | *p* value[2] |
| Margins | | | **0.006** | | | 0.331 | | | 0.456 |
|     Circumscribed / Irregular | 5 (31%) | 11 (69%) | | 12 (63%) | 7 (37%) | | 16 (42%) | 22 (58%) | |
|     Spiculated | 16 (76%) | 5 (24%) | | 14 (78%) | 4 (22%) | | 9 (53%) | 8 (47%) | |
| Shape | | | 0.680 | | | >0.999 | | | 0.871 |
|     Irregular | 17 (55%) | 14 (45%) | | 18 (69%) | 8 (31%) | | 18 (46%) | 21 (54%) | |
|     Oval / Round | 4 (67%) | 2 (33%) | | 8 (73%) | 3 (27%) | | 7 (44%) | 9 (56%) | |
| Intratumoral high signal intensity on T2 | | | 0.260 | | | 0.279 | | | 0.456 |
|     Present | 9 (69%) | 4 (31%) | | 10 (83%) | 2 (17%) | | 9 (53%) | 8 (47%) | |
|     Absent | 12 (50%) | 12 (50%) | | 16 (64%) | 9 (36%) | | 16 (42%) | 22 (58%) | |
| BES | | | 0.196 | | | 0.063 | | | 0.426 |
|     1 | 7 (50%) | 7 (50%) | | 3 (25%) | 9 (75%) | | 7 (70%) | 3 (30%) | |
|     2 | 6 (55%) | 5 (45%) | | 7 (58%) | 5 (42%) | | 11 (52%) | 10 (48%) | |
|     3 | 3 (43%) | 4 (57%) | | 1 (11%) | 8 (89%) | | 10 (59%) | 7 (41%) | |
|     4 | 5 (100%) | 0 (0%) | | 0 (0%) | 4 (100%) | | 2 (29%) | 5 (71%) | |
| Multifocality | | | **0.023** | | | 0.404 | | | 0.158 |
|     Present | 9 (90%) | 1 (10%) | | 5 (56%) | 4 (44%) | | 7 (70%) | 3 (30%) | |
|     Absent | 12 (44%) | 15 (56%) | | 21 (75%) | 7 (25%) | | 18 (40%) | 27 (60%) | |
| Background Parenchymal Enhancement | | | 0.461 | | | >0.999 | | | **0.007** |
|     Minimal/Mild | 14 (52%) | 13 (48%) | | 21 (70%) | 9 (30%) | | 24 (55%) | 20 (45%) | |
| | Moderate/Marked | 7 (70%) | 3 (30%) | | 5 (71%) | 2 (29%) | | 1 (9.1%) | 10 (91%) | |
| Associated non-mass Enhancement | | | | 0.702 | | | >0.999 | | | 0.158 |
| | Absent | 15 (54%) | 13 (46%) | | 22 (69%) | 10 (31%) | | 18 (40%) | 27 (60%) | |
| | Present | 6 (67%) | 3 (33%) | | 4 (80%) | 1 (20%) | | 7 (70%) | 3 (30%) | |
| Internal Enhancement Type | | | | **0.023** | | | >0.999 | | | **0.012** |
| | Homogeneous / Heterogeneous | 12 (44%) | 15 (56%) | | 20 (69%) | 9 (31%) | | 13 (34%) | 25 (66%) | |
| | Rim Enhancement | 9 (90%) | 1 (10%) | | 6 (75%) | 2 (25%) | | 12 (71%) | 5 (29%) | |
| Delayed Phase Enhancement | | | | 0.565 | | | >0.999 | | | 0.101 |
| | Persistent / Plateau | 6 (50%) | 6 (50%) | | 11 (73%) | 4 (27%) | | 9 (64%) | 5 (36%) | |
| | Wash-out | 15 (60%) | 10 (40%) | | 15 (68%) | 7 (32%) | | 16 (39%) | 25 (61%) | |
| Breast Composition | | | | 0.538 | | | >0.999 | | | 0.491 |
| | A-B | 10 (62%) | 6 (38%) | | 11 (69%) | 5 (31%) | | 14 (50%) | 14 (50%) | |
| | C-D | 11 (52%) | 10 (48%) | | 15 (71%) | 6 (29%) | | 11 (41%) | 16 (59%) | |
| Index Lesion MR Size | | 25 (24, 33) | 25 (21, 33) | 0.442 | 29 (22, 36) | 24 (20, 28) | 0.183 | 29 (21, 35) | 27 (22, 36) | 0.980 |
| Maximal MR Size | | 33 (24, 41) | 30 (24, 36) | 0.290 | 31 (24, 40) | 24 (22, 32) | 0.212 | 31 (23, 42) | 29 (23, 45) | 0.793 |

Table S6. Association between MRI features and Ki67 levels by breast cancer subtype (Luminal, HER2-positive, TN).

|  | Luminal | | | HER2 | | | TN | | |
|---|---|---|---|---|---|---|---|---|---|
| Characteristic | ≤ 25% KI67 N = 6[1] | > 25% KI67 N = 31[1] | p value[2] | ≤ 25% KI67 N = 8[1] | > 25% KI67 N = 29[1] | p value[2] | ≤ 25% KI67 N = 2[1] | > 25% KI67 N = 53[1] | p value[2] |
| Margins |  |  | 0.206 |  |  | 0.447 |  |  | 0.092 |
|   Circumscribed / Irregular | 1 (6.2%) | 15 (94%) |  | 3 (16%) | 16 (84%) |  | 0 (0%) | 38 (100%) |  |
|   Spiculated | 5 (24%) | 16 (76%) |  | 5 (28%) | 13 (72%) |  | 2 (12%) | 15 (88%) |  |
| Shape |  |  | 0.245 |  |  | >0.999 |  |  | >0.999 |
|   Irregular | 4 (13%) | 27 (87%) |  | 6 (23%) | 20 (77%) |  | 2 (5.1%) | 37 (95%) |  |
|   Oval / Round | 2 (33%) | 4 (67%) |  | 2 (18%) | 9 (82%) |  | 0 (0%) | 16 (100%) |  |
| Intratumoral high signal intensity on T2 |  |  | 0.643 |  |  | 0.232 |  |  | >0.999 |
|   Present | 3 (23%) | 10 (77%) |  | 1 (8.3%) | 11 (92%) |  | 0 (0%) | 17 (100%) |  |
|   Absent | 3 (12%) | 21 (88%) |  | 7 (28%) | 18 (72%) |  | 2 (5.3%) | 36 (95%) |  |
| BES |  |  | 0.181 |  |  | 0.700 |  |  | 0.172 |
|   1 | 1 (7.1%) | 13 (93%) |  | 2 (17%) | 10 (83%) |  | 0 (0%) | 10 (100%) |  |
|   2 | 1 (9.1%) | 10 (91%) |  | 4 (33%) | 8 (67%) |  | 0 (0%) | 21 (100%) |  |
|   3 | 3 (43%) | 4 (57%) |  | 1 (11%) | 8 (89%) |  | 1 (5.9%) | 16 (94%) |  |
|   4 | 1 (20%) | 4 (80%) |  | 1 25%) | 3 (75%) |  | 1 (14%) | 6 (86%) |  |
| Multifocality |  |  | 0.653 |  |  | 0.373 |  |  | **0.030** |
|   Present | 2 (20%) | 8 (80%) |  | 3 (33%) | 6 (67%) |  | 2 (20%) | 8 (80%) |  |
|   Absent | 4 (15%) | 23 (85%) |  | 5 (18%) | 23 (82%) |  | 0 (0%) | 45 (100%) |  |
| Background Parenchymal Enhancement |  |  | >0.999 |  |  | 0.156 |  |  | >0.999 |

| | | | | | | | | |
|---|---|---|---|---|---|---|---|---|
| Minimal/Mild | 5 (19%) | 22 (81%) | | 5 (17%) | 25 (83%) | | 2 (4.5%) | 42 (95%) |
| Moderate/Marked | 1 (10%) | 9 (90%) | | 3 (43%) | 4 (57%) | | 0 (0%) | 11 (100%) |
| Associated non-mass Enhancement | | | 0.620 | | | >0.999 | | | **0.030** |
| Absent | 4 (14%) | 24 (86%) | | 7 (22%) | 25 (78%) | | 0 (0%) | 45 (100%) |
| Present | 2 (22%) | 7 (78%) | | 1 (20%) | 4 (80%) | | 2 (20%) | 8 (80%) |
| Internal Enhancement Type | | | 0.313 | | | >0.999 | | | 0.092 |
| Homogeneous / Heterogeneous | 3 (11%) | 24 (89%) | | 6 (21%) | 23 (79%) | | 0 (0%) | 38 (100%) |
| Rim Enhancement | 3 (30%) | 7 (70%) | | 2 (25%) | 6 (75%) | | 2 (12%) | 15 (88%) |
| Delayed Phase Enhancement | | | 0.641 | | | >0.999 | | | >0.999 |
| Persistent / Plateau | 1 (8.3%) | 11 (92%) | | 3 (20%) | 12 (80%) | | 0 (0%) | 14 (100%) |
| Wash-out | 5 (20%) | 20 (80%) | | 5 (23%) | 17 (77%) | | 2 (4.9%) | 39 (95%) |
| Breast Composition | | | 0.680 | | | 0.705 | | | 0.491 |
| A-B | 2 (12%) | 14 (88%) | | 4 (25%) | 12 (75%) | | 2 (7.1%) | 26 (93%) |
| C-D | 4 (19%) | 17 (81%) | | 4 (19%) | 17 (81%) | | 0 (0%) | 27 (100%) |
| Index Lesion MR Size | 24 (22, 36) | 27 (22, 35) | 0.991 | 24 (21, 38) | 27 (23, 35) | 0.810 | 28 (24, 31) | 27 (21, 35) | 0.822 |
| Maximal MR Size | 35 (24, 48) | 30 (23, 40) | 0.520 | 24 (21, 38) | 30 (24, 40) | 0.438 | 47 (41, 53) | 30 (23, 42) | 0.184 |

[1] n (%)  
[2] Pearson's Chi-squared test; Fisher's exact test

**Figure S1.** Breast Edema Score illustrated by representative axial T2-weighted images. No edema is visible surrounding the tumor depicted by a curved arrow, BES1 (A). Image shows a peritumoral edema appearing as a high signal intensity surrounding an irregular mass (straight arrows), corresponding to BES 2 category (B). Prepectoral edema (BES 3) is seen as a high signal intensity is seen between the tumor and the pectoralis major muscle (thick arrow) (C). Subcutaneous edema (BES4) appears as a high-signal-intensity in the subcutaneous area (arrowheads) (D).

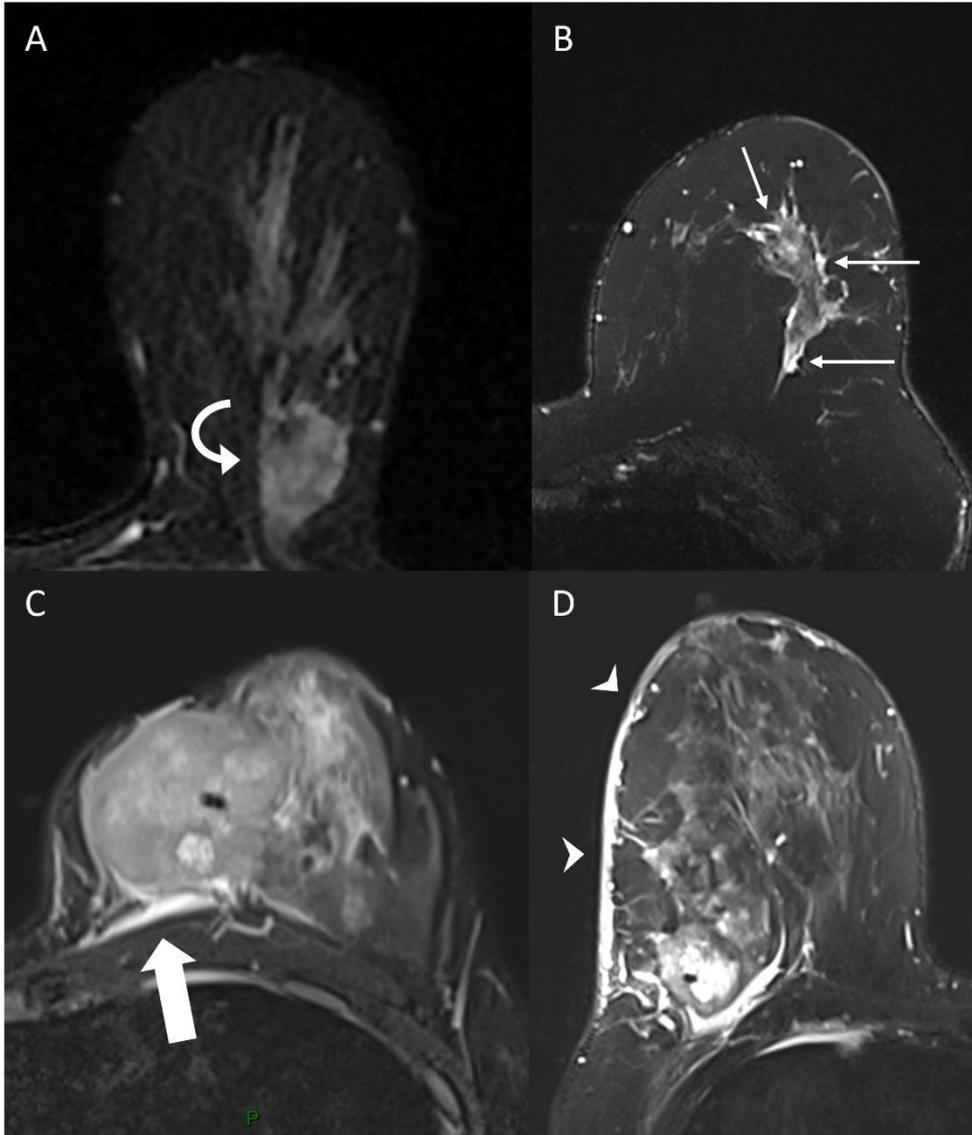

Eur Radiol (2023) Malhaire C, Selhane F, Saint-Martin MJ et al

**Figure S2.** Index Lesion MRI size distribution according to Breast Edema Score (BES) in the whole population **(A)**, in Luminal **(B)**, HER2-positive **(C)**, and TN BC **(D)**. Significant differences in MR size were shown according to BES in the whole population, as well as in luminal and TN BC. In the whole population, pairwise analyses with correction for multiple comparisons showed significant differences between BES 1, and BES 4 (*p*=0.001), BES 3 (*p*=0.002). Significant differences between BES 1 and BES 4 were also shown in Luminal BC (*p*=0.024).

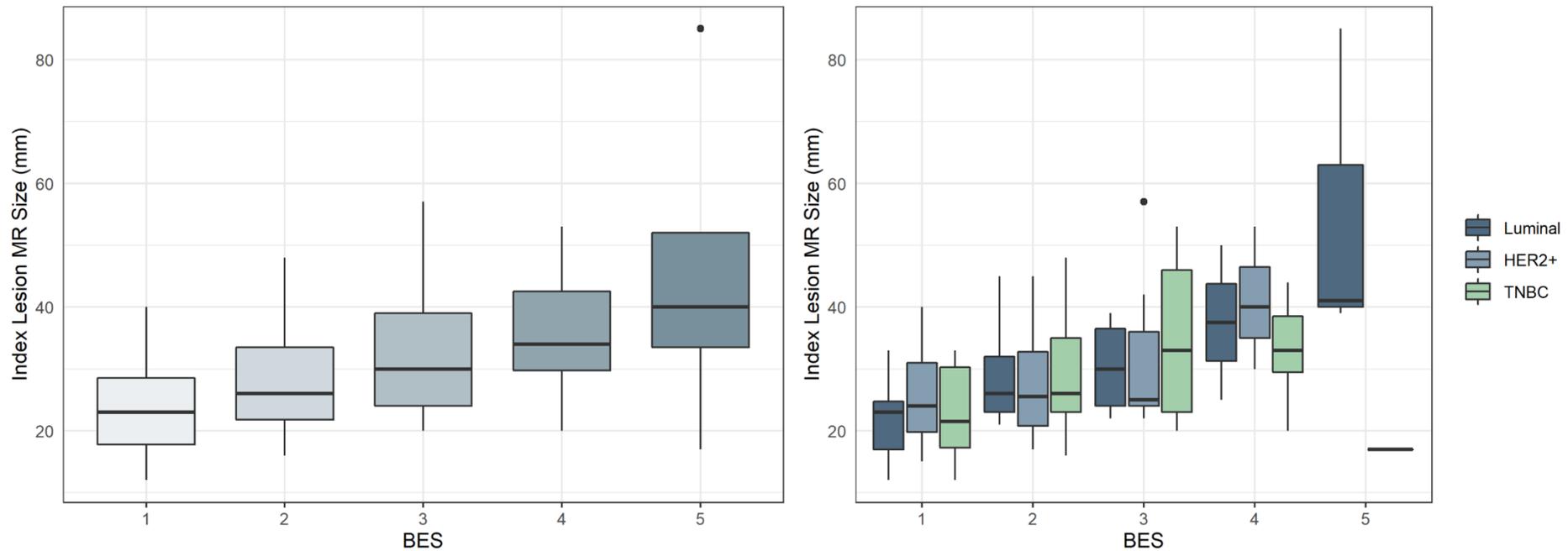

Eur Radiol (2023) Malhaire C, Selhane F, Saint-Martin MJ et al

**Figure S3.** TILs expression levels by the absence (BES 1/2/3) and presence (BES 4) of a subcutaneous edema on T2-weighted MRI

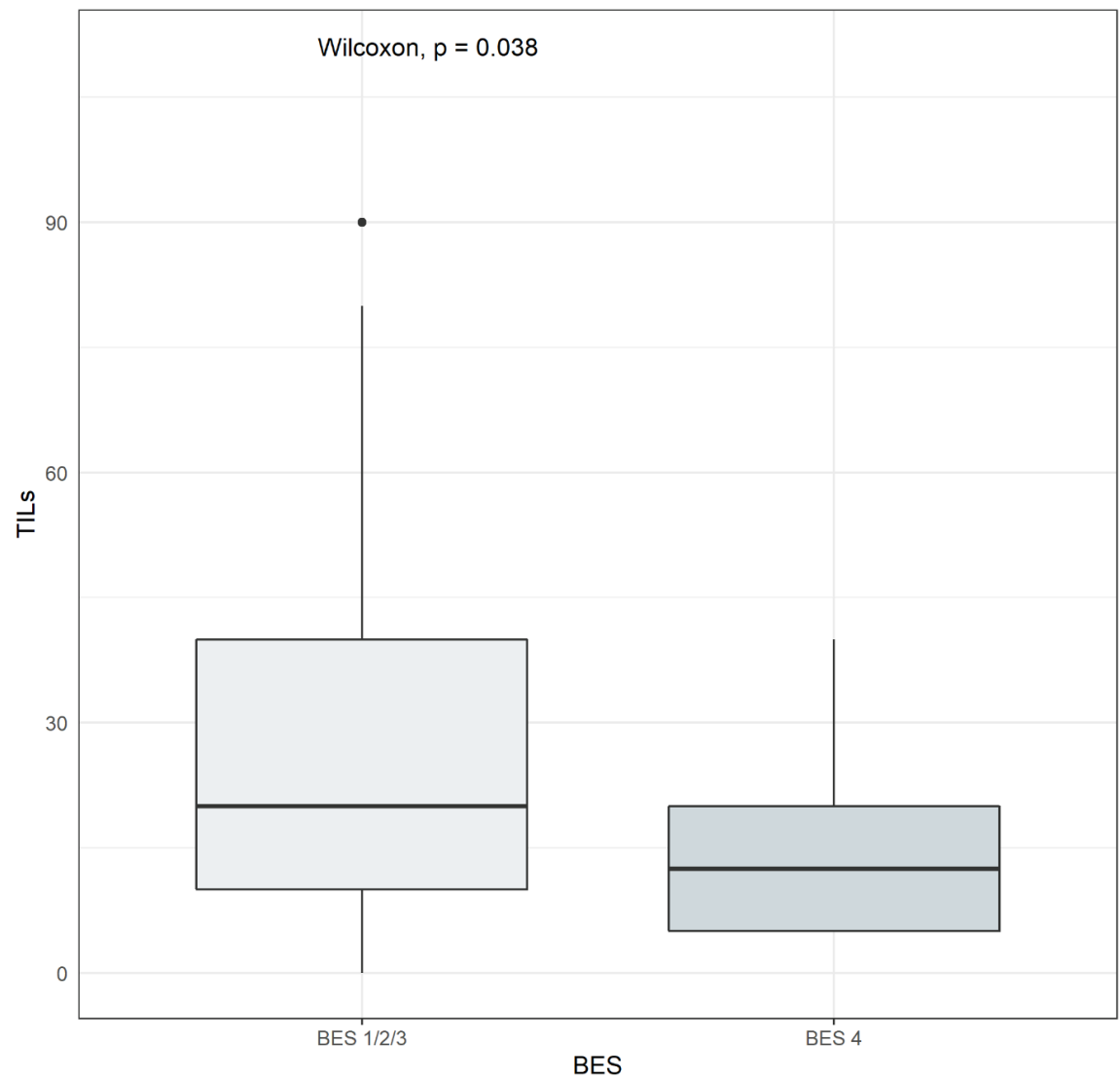

Eur Radiol (2023) Malhaire C, Selhane F, Saint-Martin MJ et al

**Figure S4.** Forest plot displaying the Odds Ratios derived from multivariable logistic regression analysis.

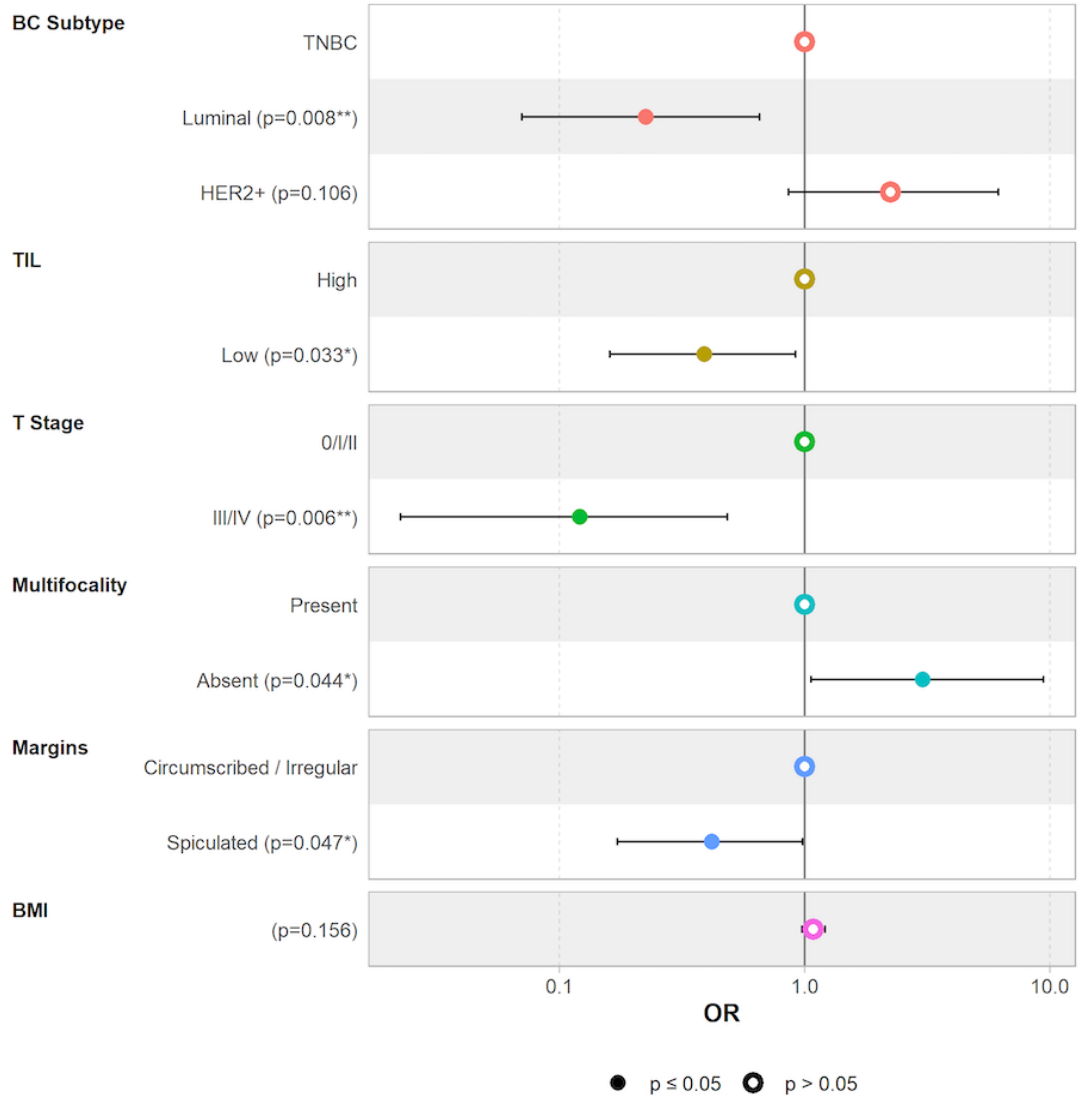

Eur Radiol (2023) Malhaire C, Selhane F, Saint-Martin MJ et al

**Figure S5.** Graph derived from multiple component analysis. Both subjects (dots) and variables (arrows) appear on the plot. Ellipses delineate the pCR (light green) and non pCR groups (dark green). The graph shows how low KI67 and low TILs levels are correlated and mostly associated with spiculated margins as the arrows from these variables are in close proximity. On the other hand, unifocality and the absence of an associated NME are closely associated with TN subtype and high TILs levels.

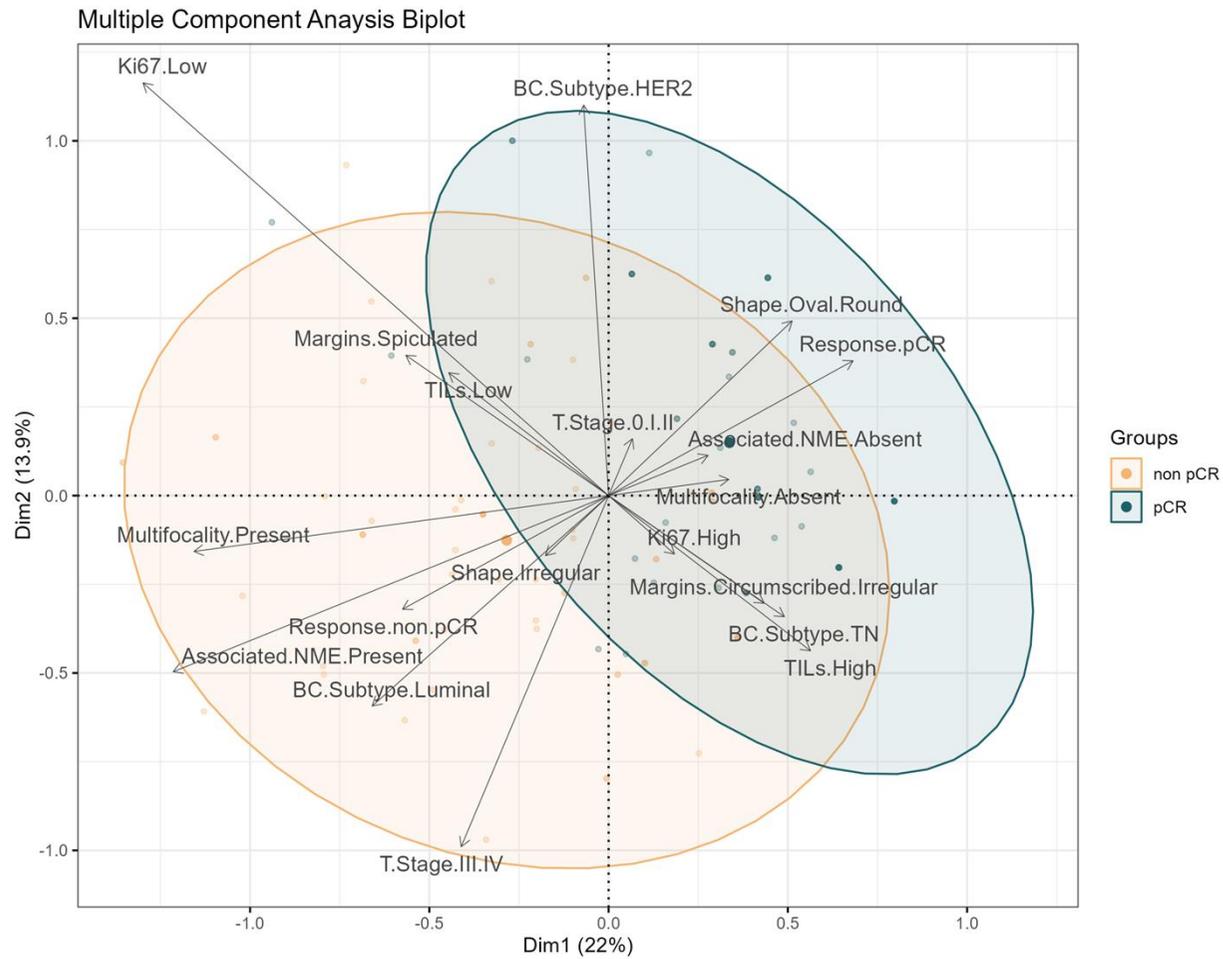

Eur Radiol (2023) Malhaire C, Selhane F, Saint-Martin MJ et al